\documentclass[12pt]{article}
\usepackage{cite} 
\usepackage{amsfonts,latexsym,amssymb}
\usepackage{graphics}
\usepackage{graphicx}
\usepackage{amssymb}
\usepackage{amsmath}
\usepackage{fancyhdr}
\usepackage{sectsty}
\usepackage{epsfig}
\usepackage{cite}
\begin{document}
\newcommand{\de}{\delta}\newcommand{\ga}{\gamma}
\newcommand{\e}{\epsilon} \newcommand{\ot}{\otimes}
\newcommand{\be}{\begin{equation}} \newcommand{\ee}{\end{equation}}
\newcommand{\ba}{\begin{array}} \newcommand{\ea}{\end{array}}
\newcommand{\beq}{\begin{equation}}\newcommand{\eeq}{\end{equation}}
\newcommand{\tmod}{{\cal T}}\newcommand{\amod}{{\cal A}}
\newcommand{\bemod}{{\cal B}}\newcommand{\cmod}{{\cal C}}
\newcommand{\dmod}{{\cal D}}\newcommand{\hmod}{{\cal H}}
\newcommand{\s}{\scriptstyle}\newcommand{\tr}{{\rm tr}}
\newcommand{\einsop}{{\bf 1}}
\def\A{{\mathcal A}}
\def\B{{\mathcal B}}
\def\C{{\mathcal C}}
\def\D{{\mathcal D}}
\def\R{\overline{R}} 
\def\doa{\downarrow}
\def\dag{\dagger}
\def\ve{\epsilon}
\def\si{\sigma}
\def\ga{\gamma}
\def\nn{\nonumber}
\def\le{\langle}
\def\re{\rangle}
\newcommand{\bra}{\langle}
\newcommand{\ket}{\rangle}
\def\lt{\left}
\def\rt{\right}
\def\dwn{\downarrow}   
\def\up{\uparrow}
\def\dag{\dagger}
\def\bea{\begin{eqnarray}}
\def\eea{\end{eqnarray}}
\def\p{\tilde{p}}
\def\q{\tilde{q}}
\def\H{\overline{H}}
\newcommand{\reff}[1]{eq.~(\ref{#1})}

\title{Emergent quantum phases in a heteronuclear molecular Bose--Einstein condensate model}
\author{
Melissa Duncan$^1$, Angela Foerster$^2$, Jon Links$^1$, \\
Eduardo Mattei$^2$, Norman Oelkers$^1$, and Arlei Prestes Tonel$^3$
\vspace{0.5cm}
\\
$^{1}$ Centre for Mathematical Physics, School of Physical Sciences, \\ 
The University of Queensland, Brisbane, 4072, Australia \\ 
$^{2}$ Instituto de F\'{\i}sica da UFRGS, \\ 
Av. Bento Gon\c{c}alves 9500, Porto Alegre, RS - Brazil \\
$^{3}$ Universidade Federal do Pampa/UFPel, \\
Rua Carlos Barbosa SN, Bag\'e, RS - Brazil
}
\vspace{0.5cm}

\maketitle

\begin{abstract}
We study a three-mode Hamiltonian modelling a heteronuclear molecular Bose--Einstein condensate. Two modes are associated with two distinguishable atomic constituents, which can combine to form a molecule represented by the third mode. 
Beginning with a semi-classical analogue of the model, we conduct an analysis to 
determine the phase space fixed points of the system. Bifurcations of the fixed points naturally 
separate the coupling parameter space into different regions. Two distinct scenarios are found, dependent
on whether the imbalance between the number operators for the atomic modes is zero or non-zero.
This result suggests the ground-state properties of the model exhibit an unusual sensitivity on the atomic imbalance. We then test this finding for the quantum mechanical model. Specifically we use Bethe ansatz methods, ground-state expectation values, the character of the quantum dynamics, and ground-state wavefunction overlaps to clarify the nature of the ground-state phases. The character of the transition is smoothed due to quantum fluctuations, but we may nonetheless identify the emergence of a quantum phase boundary in the limit of 
zero atomic imbalance. 
\end{abstract}

PACS: 02.30.Ik, 03.65.Sq, 03.75.Nt

\vfil\eject
%%%%%%%%%%%%%%%%

\section{Introduction}

The achievement of producing Bose--Einstein condenstates with ultracold dilute gases of atoms has seen a wealth of theoretical and experimental activity. One enticing prospect of Bose-Einstein condensates is that they may allow for a better understanding of the interface between classical and quantum mechnics, through the possibility of macroscopic 
Schr\"odinger cat states \cite{zoller} and macroscopic quantum tunneling \cite{leggett}. Another intriguing field of study is the chemistry of Bose-Einstein condensates, where the atomic constituents may form molecules through Feshbach resonances 
\cite{feshbach} or photoassociation \cite{photo}. A novel feature of a molecular Bose--Einstein condensate is that the atomic and molecular states can exist as a superposition 
\cite{donley}, providing a chemical analogue of a Schr\"odinger cat state. In cases where the molecules are heteronuclear, the presence of a permanent electric dipole moment also opens the possibility for manipulating the condensate through electrostatic forces \cite{stwalley}. 

Since systems of Bose--Einstein condensates exist at ultracold temperatures, it is to be expected that significant insights into their behaviour can be obtained from studying their ground-state properties.  
From a general theoretical perspective there has been substantial progress in the understanding of quantum (i.e. ground-state) phases in many-body quantum systems, due largely to a cross fertilisation of ideas between the condensed matter theory and the quantum information theory communities. Much of this study has explored the relationship between entanglement and quantum criticality \cite{on,oaff,vlrk}. 
However other characterisations of quantum criticality have been sought too \cite{vbkos,hamma}. Recently the notion of wavefunction overlaps (also known as the {\it fidelity}), which is again common in quantum information theory, has  been applied to the study of quantum phase transitions \cite{zanardi,huanandjp}.  An advantage of this approach is its universality, as it can be applied to any system independent of the choice of decomposition into subsystems.      
    
With the above points in mind here we analyse a simple, yet non-trivial, three-mode model describing a heteronuclear molecular condensate. Two modes are associated with two distinguishable atomic constituents, which can combine to form a molecule represented by the third mode. Besides the interaction describing the interconversion of atoms and molecules, the Hamiltonian contains terms which are linear in the mode number operators (corresponding to external fields) and terms which are second-order in the mode number operators (corresponding to scattering interactions between atoms and molecules). 
We mainly concern ourselves with the ground-state properties of the model, with the aim of identifying the ground-state phases. We avoid taking the thermodynamic limit and restrict our analysis to finite systems, for reasons which will be discussed later. This in turn presents challenges in rigourously identifying quantum phases, since for finite systems there are no singularities in physical quantities such as the ground state energy and its derivatives. However several recent works have addressed the issues of quantum phases in finite systems \cite{iz,dhl,adgv,leviatan}. We mention that traditional techniques of renormalisation group methods are not applicable to the model under consideration, due to the low number of degrees of freedom. Neither is the concept of symmetry breaking, as the model does not admit global symmetries, nor long-range order, as the model is in essence zero-dimensional.   

We start our analysis with a semi-classical treatment, following the approach of \cite{fitonel}. Since the model with which we are dealing is integrable, the semi-classical many-body system can be reduced to a problem with a single degree of freedom. We study the phase space of this system, in particular determining the fixed points. It is found that for certain coupling parameters bifurcations of the fixed points occur, and we can determine a parameter space diagram which classifies the  fixed points. An unexpected result is that the boundaries between the regions in parameter space are extremely sensitive on whether the number of constituent atoms is equal or not. Specifically, when the number of constituent atoms is equal (i.e. the {\it atomic imbalance} is zero) there is a spontaneous appearance of additional boundaries in the parameter space, some of which can be identified with bifurcations of the global minimum of the classical Hamiltonian. 

We next investigate the extent to which the classical behaviour influences the ground-state properties of the quantum system. Our first goal in the full quantum analysis is to derive an exact Bethe ansatz solution for the model. We use the Bethe ansatz solution to map the spectrum of the Hamiltonian into that of a one-body 
Schr\"odinger equation in one-dimensional. An advantage of this method is that it allows for an analysis of the finite system, following the ideas of \cite{dhl}, as the mapping to the one-body Schr\"odinger equation  is not dependent on taking the thermodynamic limit of the original many-body system. The results of the analysis of the associated Schr\"odinger equation are in general agreement with the results obtained from the semi-classical treatment, supporting the picture of an additional phase boundary when the atomic imbalance is zero. 
However, due to quantum fluctuations, the emergence of the phase boundary is smooth rather than spontaneous. This property is apparent from a study of ground-state expectation values and quantum dynamics.    

In order to simply characterise the ground-state phases for the finite system, we finally define the notion of a {\it quantum phase pre-transition} in terms of wavefunction overlaps.  Specifically, a quantum phase pre-transition is identified with each coupling for which the incremental ground-state wavefunction overlap is a local minimum. We numerically calculate these for several cases and discuss these results in relation to the semi-classical and quantum analyses which have been described above. The results confirm the emergence of a quantum phase boundary in the limit of zero atomic imbalance.

\section{The model}
We consider a general three-mode Hamiltonian describing a heteronuclear molecular Bose--Einstein condensate with two distinct 
species of atoms, labelled by $a$ and $b$, which can combine to produce a molecule labelled by $c$. We introduce canonical creation and annihilation operators $\{a,\,b,\,c,\,a^{\dagger},\,b^{\dagger},\,c^{\dagger}\}$ satisfying the usual commutation relations $[a,\,a^\dagger]=I$ etc., which represent 
the three degrees of freedom in the model. The Hamiltonian reads \cite{jzrg}
\begin{eqnarray}
H&=&U_{aa}N_a^2 + U_{bb}N_b^2 +U_{cc}N_c^2+U_{ab}N_aN_b+U_{ac}N_aN_c+U_{bc}N_bN_c \nonumber \\
 &+& \mu_aN_a +\mu_bN_b+\mu_cN_c + \Omega(a^{\dag}b^{\dag}c +c^{\dag}ba).
\label{ham}
\end{eqnarray}
The parameters $U_{ij}$ describe S-wave scattering,
$\mu_i$ are external potentials and $\Omega$ is the amplitude for interconversion of atoms and molecules.
We remark that in the limit
$U_{aa}=U_{bb}=U_{cc}=U_{ab}=U_{ac}=U_{bc}=0$, equation  
(\ref{ham}) is the Hamiltonian studied in \cite{wb,wt}
in the context of quantum optics. In the latter stages of the manuscript we will 
study this limiting case in some detail.

The Hamiltonian acts on the Fock space spanned by the (unnormalised) vectors
\begin{equation}
 \left|n_a;n_b;n_c\right> 
={(a^\dagger)^{n_a}(b^\dagger)^{n_b}(c^\dagger)^{n_c}}\left|0\right>
\label{states}
\end{equation}
where $\left|0\right>$ is the Fock vacuum. We then have  
$$N_a \left|n_a;n_b;n_c\right> = n_a \left|n_a;n_b;n_c\right> $$ 
etc., where $N_a=a^{\dagger}a$, $N_b=b^{\dagger}b$ and $N_c=c^{\dagger}c$.  
The Hamiltonian commutes 
with $J=N_a-N_b$ and the total atom 
number $N=N_a+N_b+2N_c$. We refer to $J$ as the atomic imbalance and introduce $k=J/N,\,k\in[-1,1]$ 
as the fractional atomic imbalance.
As there are three degrees of freedom and three conserved operators, the system is integrable. This fact will allow us to analyse the model in some depth. Below we begin with a semi-classical analogue of the model, and determine the fixed points of the system.

\section{Semi-classical analysis} \label{sca}

Let $N_j,\,\phi_j,\,j=a,\,b,\,c$ be
quantum variables satisfying the canonical relations 
$$[\phi_j,\,\phi_k]=[N_j,\,N_k]=0,~~~~~[N_j,\,\phi_k]=i\delta_{jk}I.$$  
We make a change of variables from the operators $\{j,\,j^\dagger |\,j=a,\,b,\,c\}$ to a number-phase representation via
$$j=\exp(i\phi_j)\sqrt{N_j} \;\;\;\;\;\;\;j=a,\,b,\,c$$ 
such that the canonical commutation relations are preserved.  
We now make a further change of variables
$$ z=\frac{1}{N}(N_a+N_b-2N_c),$$
$$\phi=\frac{N}{4}(\phi_a+\phi_b-\phi_c),$$ 
such that $z$ and $\phi$ are canonically conjugate variables; i.e.
$$[z,\,\phi]=iI. $$   
For large $N$ we can now approximate the (rescaled) Hamiltonian by 
\bea
H=\lambda z^2 + 2(\alpha-\lambda)z +\lambda -2\alpha + \beta   
+\sqrt{2(1-z)(z+c_+)(z+c_-)} \cos\left(\frac{4\phi}{N}\right)
\label{ham2}
\eea
with
\begin{eqnarray*}
 \lambda &=& \frac{\sqrt{2N}}{\Omega}\left(\frac{U_{aa}}{4}+
\frac{U_{bb}}{4}+\frac{U_{cc}}{4}+ \frac{U_{ab}}{4}-\frac{U_{ac}}{4}-\frac{U_{bc}}{4}\right)  \\
\alpha &=&\frac{\sqrt{2N}}{\Omega}\left( \frac{1+k}{2}U_{aa}
+\frac{1-k}{2}U_{bb}+\frac{1}{2}U_{ab} -\frac{1+k}{4}U_{ac}-\frac{1-k}{4}U_{bc}+\frac{1}{2N}(\mu_a+\mu_b-\mu_c)\right)   \\
\beta &=& \frac{\sqrt{2N}}{\Omega}\left((1+k)^2U_{aa}+(1-k)^2U_{bb}+(1-k^2)U_{ab}+\frac{2}{N}((1+k)\mu_a+(1-k)\mu_b) \right)   
\end{eqnarray*}
where $c_{\pm}=1\pm 2k $. 
Since $N$ and $k$ are conserved, we treat them as constant. 

We now regard (\ref{ham2}) as a classical Hamiltonian and 
investigate the fixed points of the system. The first step is to 
derive Hamilton's equations of motion yielding  
\begin{eqnarray*} 
\frac{dz}{dt}=\frac{\partial H}{\partial \phi}&=& - \frac{4}{N}\sqrt{2(1-z)(z+c_+)(z+c_-)}    \sin\left(\frac{4\phi}{N}\right),  \label{de1} \\
-\frac{d\phi}{dt}=\frac{\partial H}{\partial z} &=&2\lambda z +2\alpha-2\lambda +\frac{(1-z)(2z+2)-(z+c_+)(z+c_-)}{\sqrt{2(1-z)(z+c_+)(z+c_-)}} \cos\left(\frac{4\phi}{N}\right).
\label{de2}  
\end{eqnarray*}
The fixed points of the system are determined by the condition
\begin{equation}
\frac{\partial H}{\partial \phi}=\frac{\partial  H}{\partial z}=0. 
\label{fixed}
\end{equation}
Due to periodicity of the solutions, below we restrict to $\phi\in[0,\,N\pi/2)$. It is necessary to treat the cases of 
$k\neq 0$ and $k=0$ separately, and without loss of generality we assume $k\geq 0$.

\subsection{Case I: $k\neq 0$}

Define the functions 
\begin{eqnarray}
f(z)&=& \lambda z+\alpha-\lambda \label{f} \\
g(z)&=&  \frac{(z-1)(2z+2)+ (z+c_+)(z+c_-)}{2\sqrt{2(1-z)(z+c_+)(z+c_-)}} \label{g} 
\end{eqnarray}
Note that  the domain  of $g(z)$ is $z \in [2k-1,1]$, and $g(z)$ is divergent at $z=2k-1$ and $z=1$. 
For $k\neq 0$, we then have the following classification of solutions for (\ref{fixed}):
\begin{itemize} 
\item $\phi=0$, and $z$ is a solution of 
\begin{equation}
f(z)=g(z)  
\label{sol2}
\end{equation}
which can admit one, two or three solutions. 

\item $\phi={N\pi}/4$, and $z$ is a solution of 
\begin{equation}
f(z)=-g(z)
\label{sol1}
\end{equation}
which can admit one, two or three solutions. 
\end{itemize} 

\begin{figure}[ht]
\begin{center}
\epsfig{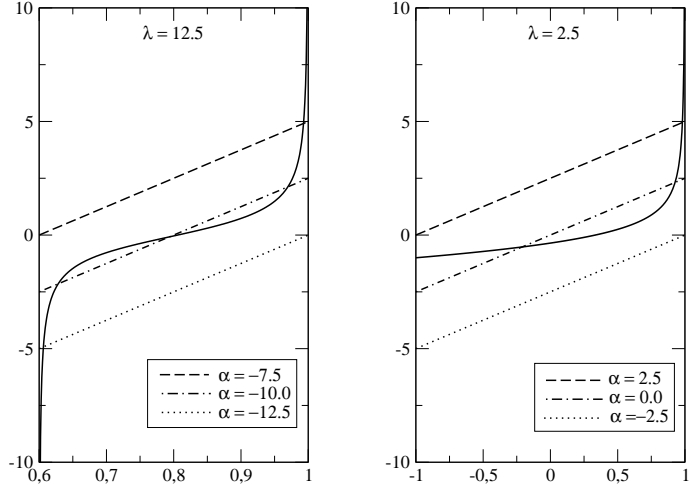}  \\          
\end{center}
\caption{On the left, graphical solution of (\ref{sol2}) with $k=0.8$. Depending on the values of $\lambda$ and $\alpha$, there may be one, two or three solutions. 
On the right, graphical solution of (\ref{ssol2}) with $k=0$. Depending on the values of $\lambda$ and $\alpha$, there may be zero, one or two solutions.} 
\label{curve2}
\vspace{1.00cm}
\end{figure}

\begin{figure}[ht]
\begin{center}
\epsfig{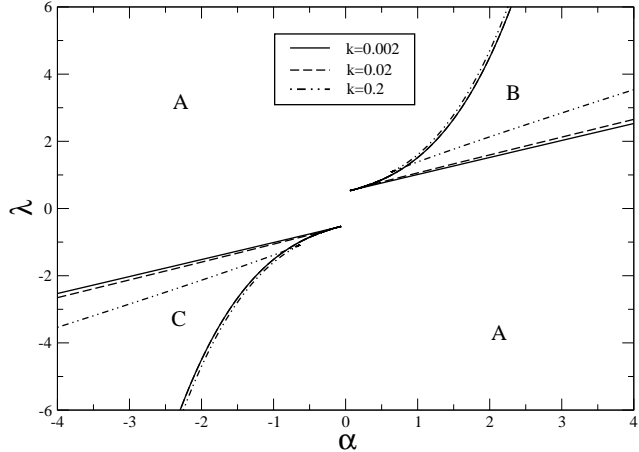}
\end{center}
\caption{Parameter space diagram identifying the different types of solutions for 
equation (\ref{fixed}), for  
$k=0.002,\,0.02,\,0.2$. In each case  
the diagram is divided into regions  
$A$ (one solution  for $z$ when $\phi = 0$ and one solution when $\phi=N\pi/4$ ), 
$B$ (three solutions for $z$ when $\phi = 0$ and one solution when $\phi=N\pi/4$) and 
$C$ (one solution for $z$ when $\phi = 0$ and three solutions when $\phi=N\pi/4$).
The boundary separating the regions is given by solutions to the equations (\ref{boundarya},\ref{boundaryb}). 
}
\label{curvekdiff}
%\vspace{1.00cm}
\end{figure}

A graphical representation of possible types of solutions for $\theta=0$ is given in Fig. \ref{curve2}. 
From the equations (\ref{sol2}, \ref{sol1}) we can determine there are fixed point bifurcations for 
certain choices of the coupling parameters. These bifurcations allow us to divide the coupling parameter space into 
different regions. To construct this diagram, we observe that bifurcations occur when $f$ is the tangent line to $g_{\pm}$; i.e. for values of $\lambda,\,\alpha$ such that 
\begin{eqnarray}
\lambda&=&\pm\left.\frac{dg}{dz}\right|_{z_0} \label{boundarya} \\
f(z_0)&=& \pm g(z_0) \label{boundaryb}
%\label{f=g}
\end{eqnarray}
for some $z_0$. This requirement determines the boundaries in parameter space, which are depicted in Fig. 
\ref{curvekdiff} 

\subsection{ Case II: $k=0$}

Next we consider the case $k=0$ for which the function $g(z)$ has substantially different properties. Setting 
$c_+=c_-=1$ into (\ref{g}), we find that $g(z)$ reduces to  
\begin{equation}
g(z)= \frac{1-3z}{2\sqrt{2(1-z)}}
\end{equation} 
Here we observe that $g(z)$ is divergent at $z=1$, but finite at $z=-1$. This property affects the types of solutions for (\ref{fixed}). Specifically, we now have the following classifications of solutions for $k=0$:

\begin{itemize} 
\item $\phi=0$, and $z$ is a solution of 
\begin{equation}
f(z)=g(z)  
\label{ssol2}
\end{equation}
which can admit zero, one or two solutions. 

\item $\phi={N\pi}/4$, and $z$ is a solution of 
\begin{equation}
f(z)=-g(z)
\label{ssol1}
\end{equation}
which can admit zero, one or two  solutions.

\item $z=-1$ and $\phi$ is a solution of 
\begin{equation}
\cos\left(\frac{4\phi}{N}\right)=-2\lambda+{\alpha} 
\label{ssol3}
\end{equation}
which can admit zero, one or two solutions. 
\end{itemize} 

\begin{figure}[h]
\begin{center}
\epsfig{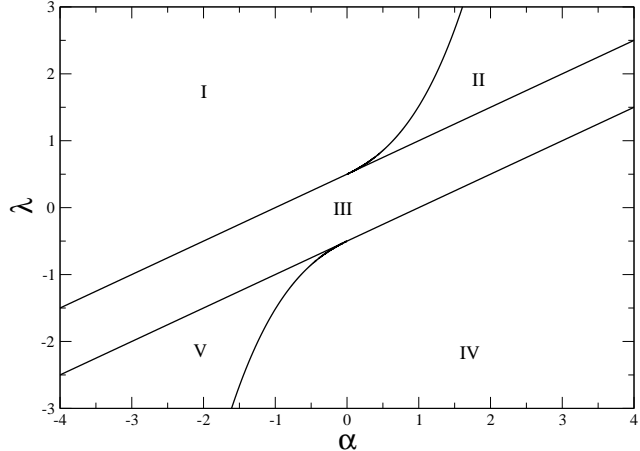}          
\end{center}
\caption{Parameter space diagram identifying the different types of solutions for 
equation (\ref{fixed}) when $k=0$. 
In region I there is no solution for $z$ when $\phi = 0$, and one solution for 
$z$ when $\phi = {N\pi}/{4}$. In region II there are two solutions for $z$ when $\phi = 0$, and one solution for 
$z$ when $\phi = {N\pi}/{4}$. In region III exists one solution for $z$ when $\phi = 0$,  one solution for 
$z$ when $\phi = {N\pi}/{4}$, and two solutions for $\phi$ when $z=-1$. In region IV there is one solution for $z$ 
when $\phi = 0$,  and no solution for $z$ when  $\phi = {N\pi}/{4}$. In region V there is one solution for $z$ when $\phi=0$, and two solutions for $z$ when  $\phi = {N\pi}/{4}$. The boundary separating regions II and III 
is given by $\lambda=(\alpha+1)/2$, while the equation $\lambda =(\alpha-1)/2$ separates the regions III and IV. 
The boundary between regions I and II has been obtained numerically.   
}
% The full line is 
% for $\phi=0$ while  dashed line is for $\phi=N\pi/4$. Here, we are using $\lambda > 0$.} 
\label{parametro1}
\end{figure}

A graphical representation of possible types of solutions for $\theta=0$ is given in Fig. \ref{curve2}.
Because $g(-1)$ is finite, for this case there can be either zero, one or two solutions. As in the $k\neq 0$ case, we can determine the region boundaries in parameter space from equations (\ref{boundarya},\ref{boundaryb}). Moreover, because of the existence of solutions of the form given by (\ref{ssol3}) for $k=0$, which do not have an analogue for $k\neq 0$, we see the appearance of new boundaries given by the conditions $\lambda=(\alpha\pm 1)/2$ for all values of $\alpha$. 
The boundaries in parameter space are depicted in Fig. \ref{parametro1}. 

\begin{figure}[h]
\begin{center}
\begin{tabular}{cccc}
\epsfig{file=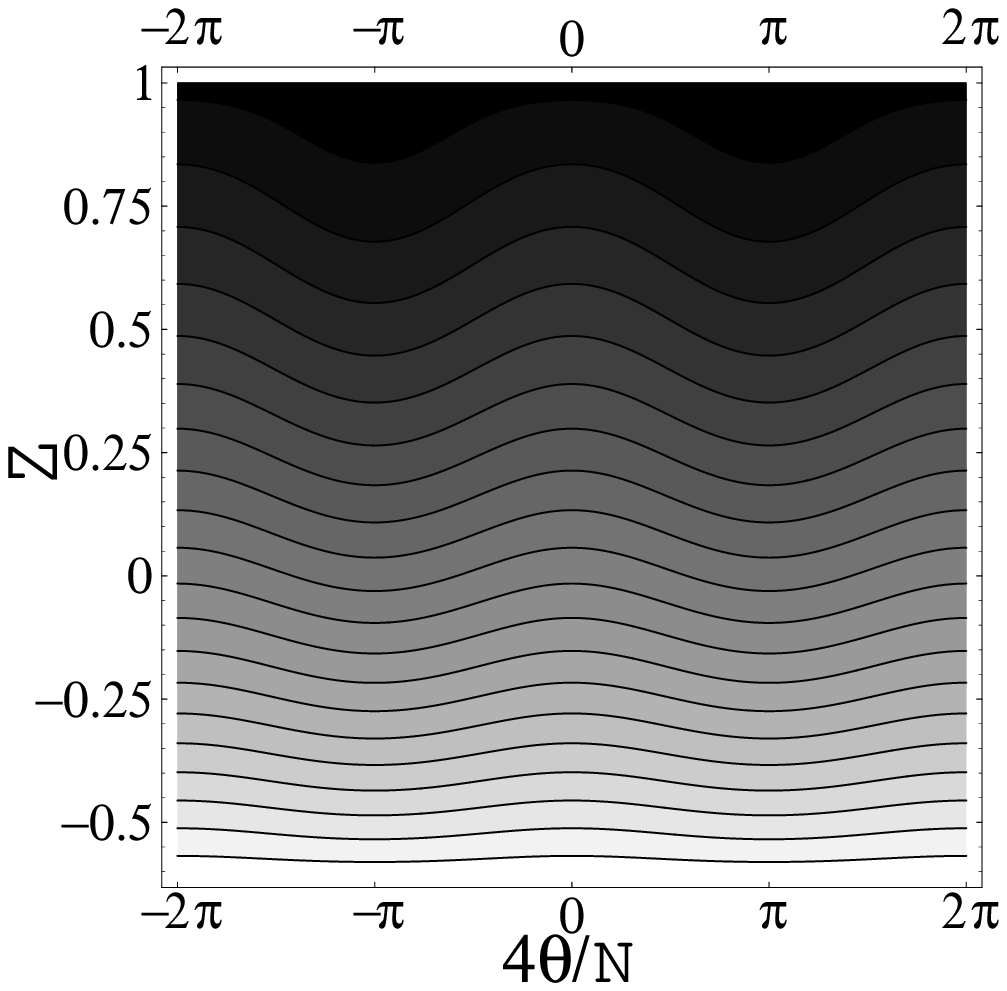,width=3.5cm,height=5cm,angle=0}&            
\epsfig{file=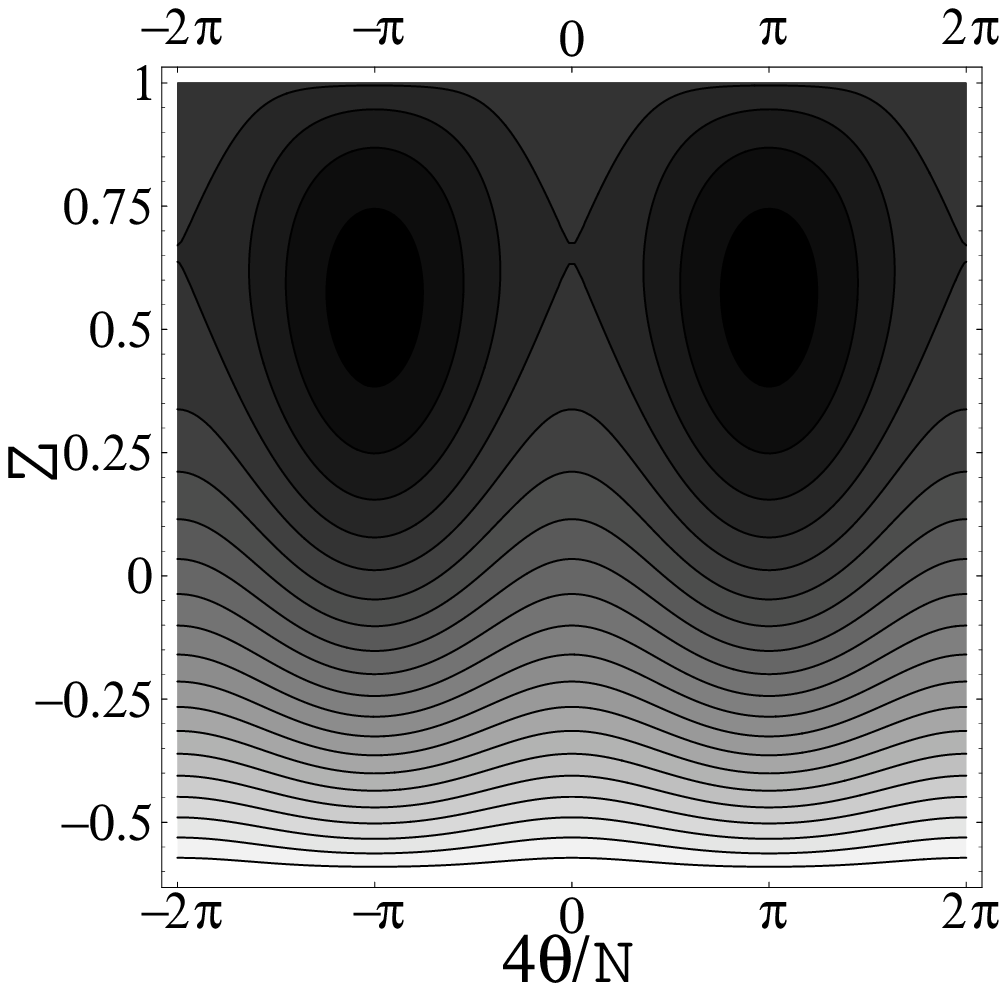,width=3.5cm,height=5cm,angle=0} &
\epsfig{file=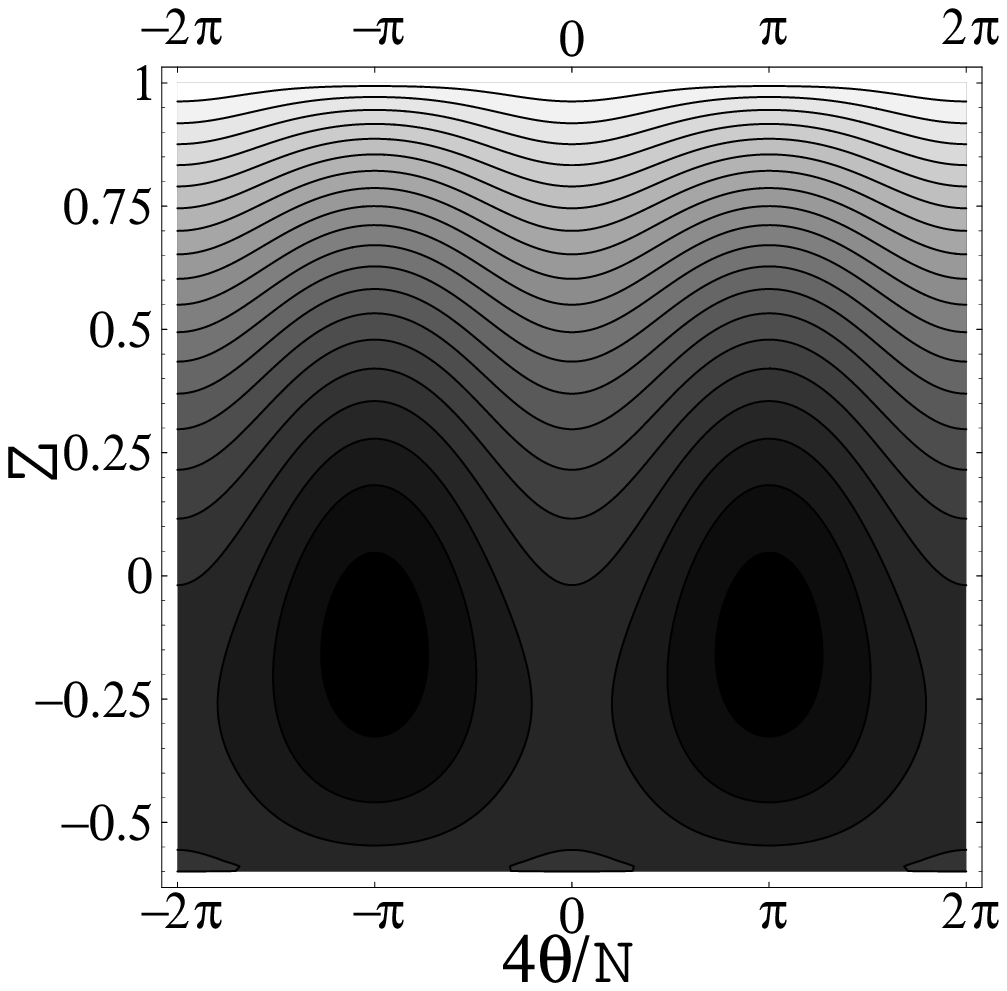,width=3.5cm,height=5cm,angle=0}&            
\epsfig{file=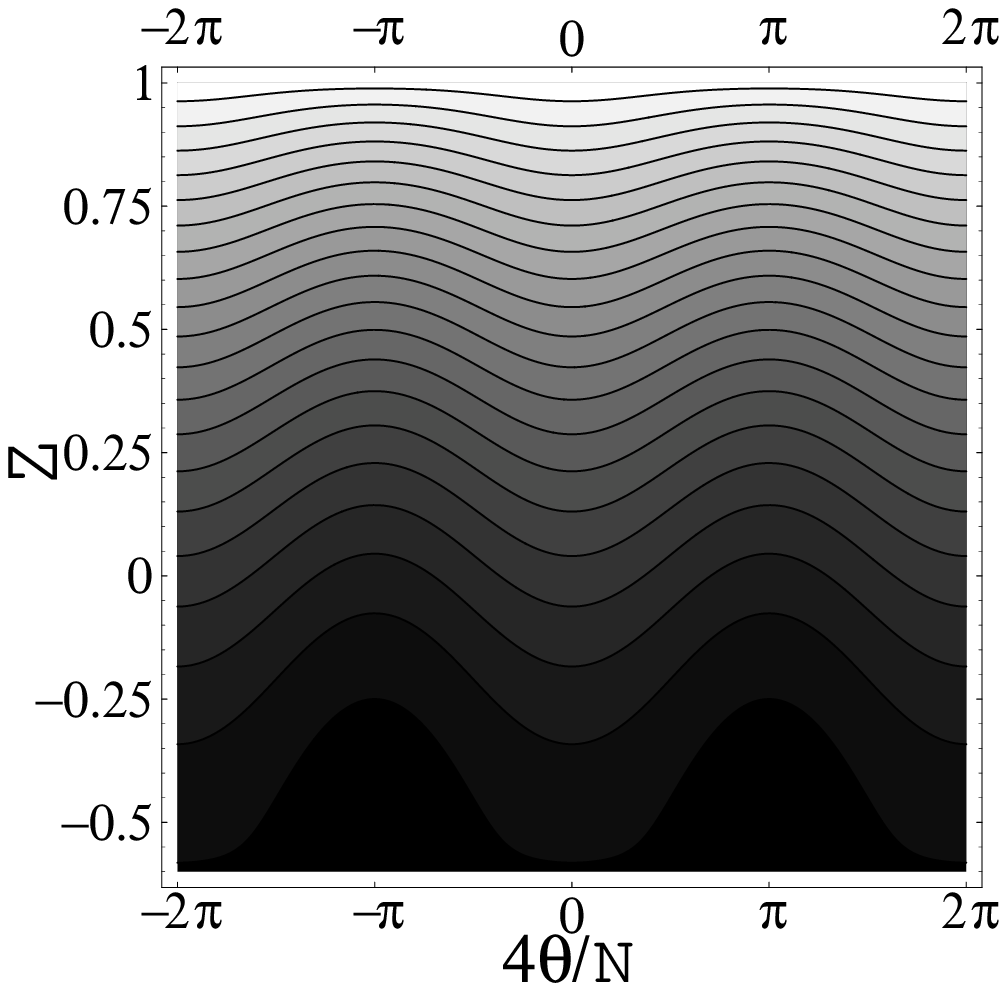,width=3.5cm,height=5cm,angle=0} \\
 (a)  & (b)    & (c) &  (d)\\     
\end{tabular}
\end{center}
\caption{Level curves of the Hamiltonian (\ref{ham2}) for $k=0.2$, where the dark regions indicate lower values than the light regions. Figures (a) and (d) correspond to  region A while
Figures (b), and (c) correspond to region B. The parameter values are: (a) $\lambda=10,\,\alpha=-5$; (b) $\lambda=10,\,\alpha=4$; 
(c) $\lambda=10,\,\alpha=12$ and (d) $\lambda=10,\,\alpha=16$. 
In region A, there is a maximal point at $\phi=0$ and minima at 
$4\phi/N=\pm\pi$. Two additional  fixed points, a saddle and a maximum,  occur in region 
B at $\phi=0$.
} 
\label{levelcurve2}
\end{figure}

\begin{figure}[ht]
\begin{center}
\begin{tabular}{cccc}
\epsfig{file=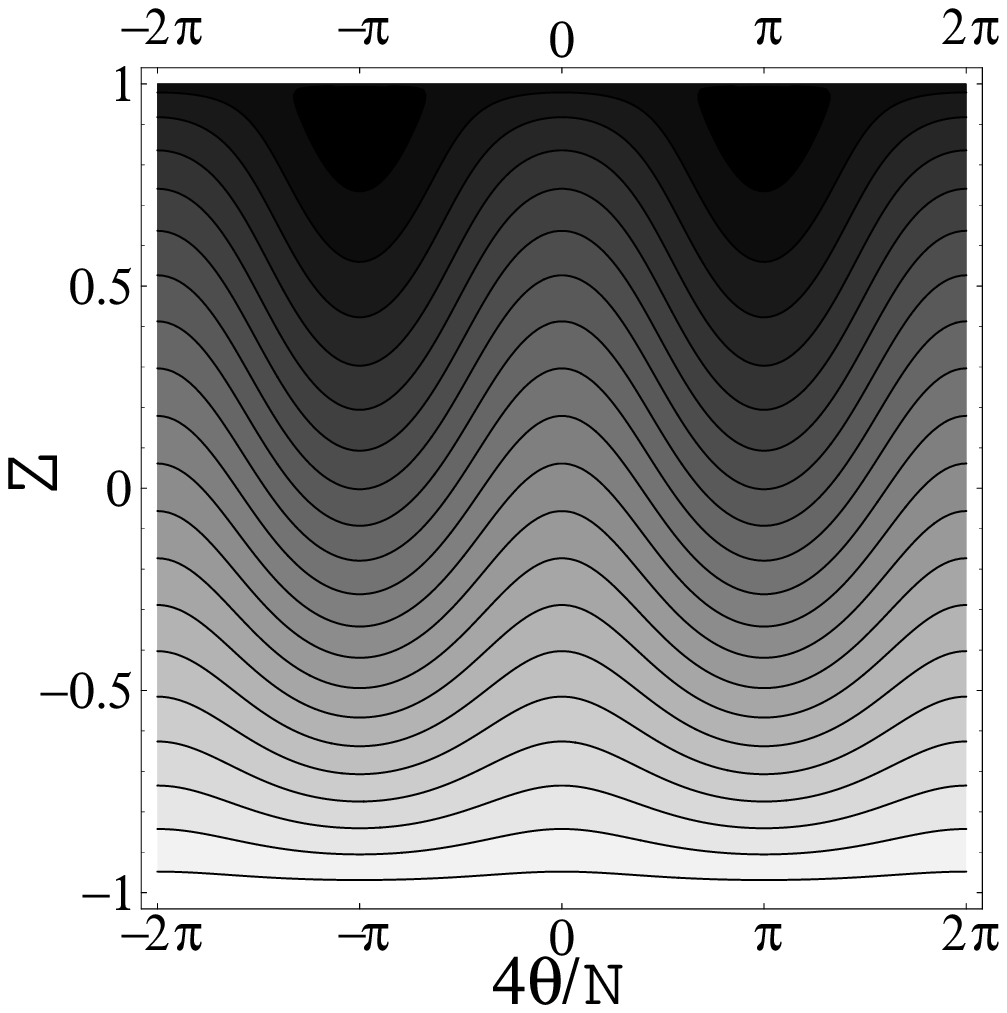,width=3.5cm,height=5cm,angle=0}&            
\epsfig{file=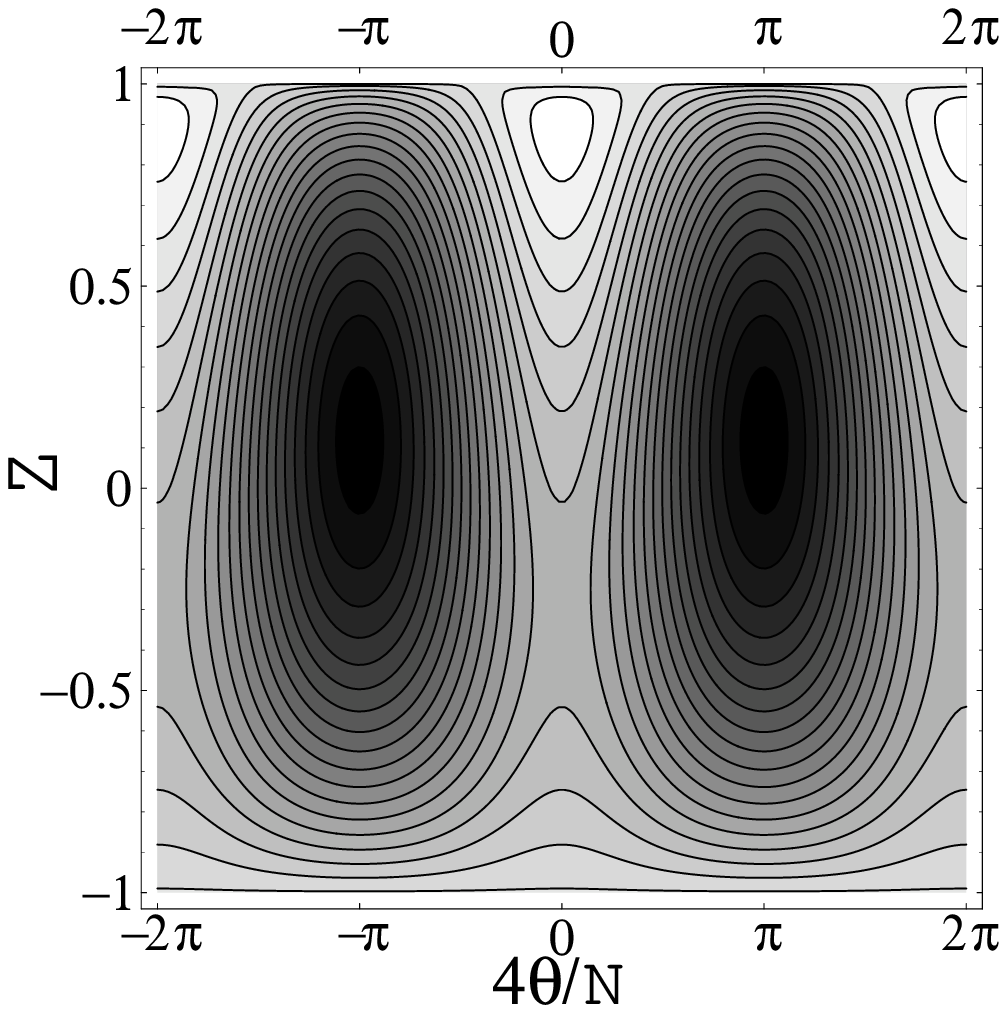,width=3.5cm,height=5cm,angle=0}&
\epsfig{file=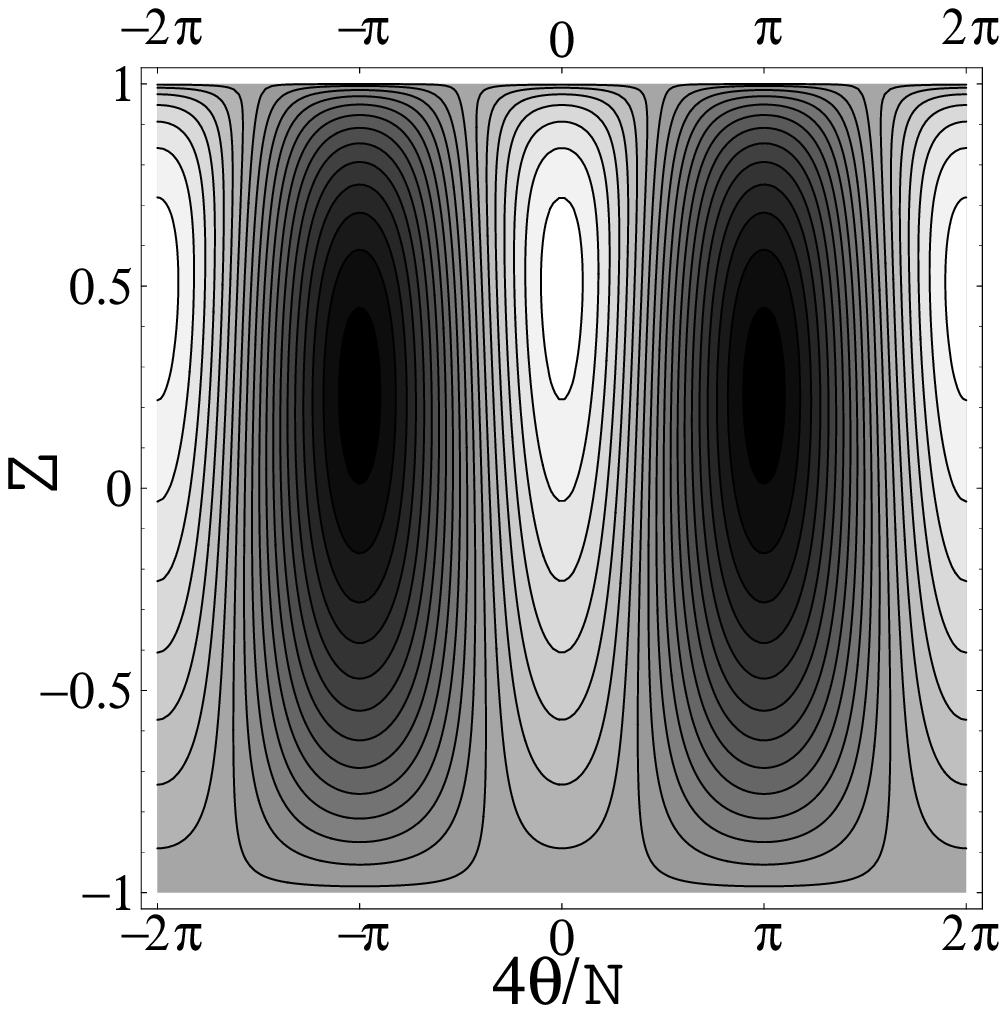,width=3.5cm,height=5cm,angle=0}&            
\epsfig{file=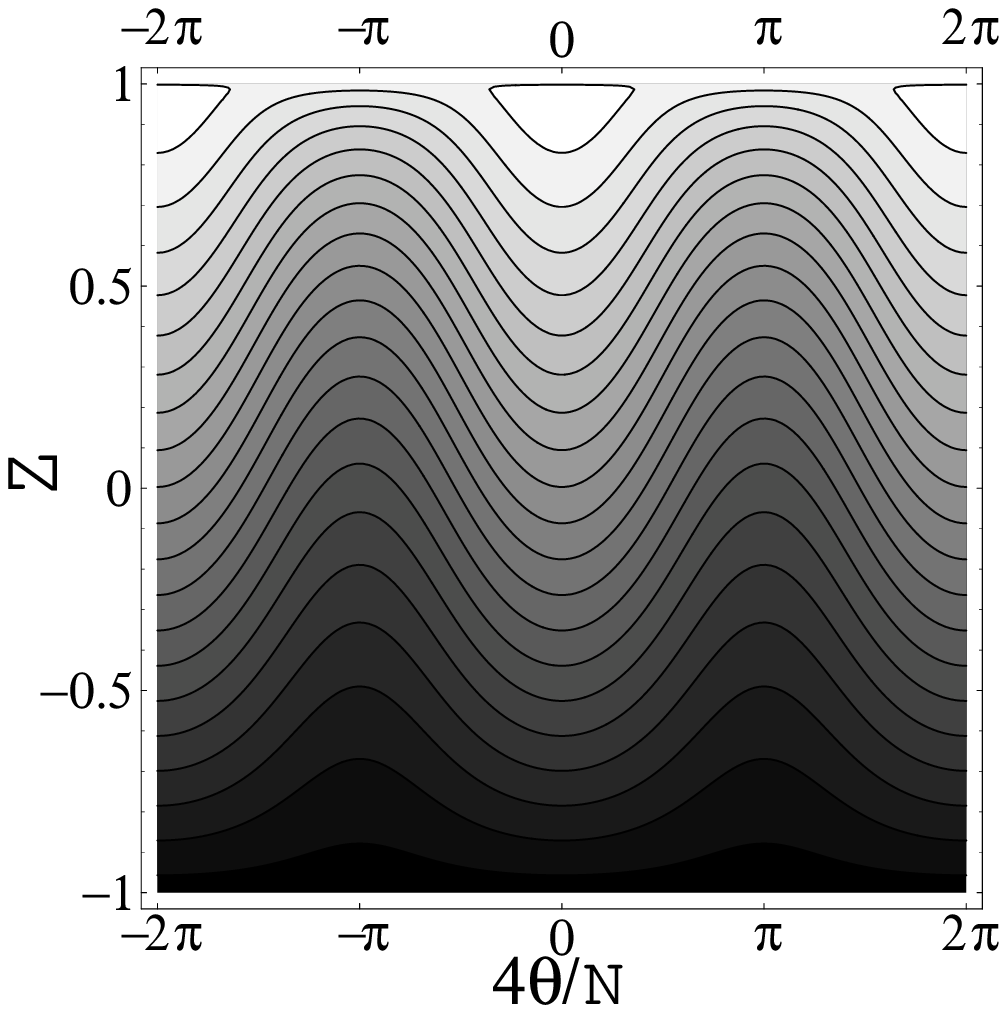,width=3.5cm,height=5cm,angle=0} \\
  (I)& (II)    &  (III)& (IV)    \\
\end{tabular}
\end{center}
\caption{Level curves of the Hamiltonian (\ref{ham2}) for $k=0$, showing the typical behaviour for regions I, II, 
III and IV. The dark regions indicate lower values than the light regions.
The parameter values are $\lambda=1.0,\,\alpha=-2.0$ for region I, $\lambda=2.0,\,\alpha=2.0$ for region II,  $\lambda=0.5,\,\alpha=0.5$ for region III 
and  $\lambda=0.5,\,\alpha=3.0$ for region IV. 
In region I there are local minima for $4\phi/N=\pm\pi$. 
Besides the minima at $4\phi/N=\pm\pi$,
two additional fixed points (a maximum and a saddle point) 
are apparent in region II occurring at $\phi=0$. In region III there are minima at $4\phi/N=\pm\pi$ 
and for $\phi=0$ just one fixed point, a maximum. There are also saddle points for when $z=-1$. 
In region IV just one fixed point, a maximum, occurs for $\phi=0$, which always has $z<1$. In contrast the global 
minimum occurs for $z=-1$. 
} 
\label{levelcurve1}
\end{figure}

To help visualise the classical dynamics, it is useful to plot the level curves of the Hamiltonian (\ref{ham2}). 
Since the fixed point bifurcations change the topology of the level curves, qualitative differences can be observed 
between each of the regions. The results are shown respectively in Fig. \ref{levelcurve2} for $k=0.2$ and 
Fig. \ref{levelcurve1} for $k=0$, where 
for clarity we show $4\phi/N\in[-2\pi,\,2\pi]$.

Hereafter we will focus most attention on the case where $\lambda=0$, so the model has one effective coupling parameter, $\alpha$. From Figs. 
\ref{curvekdiff}, \ref{parametro1}, it can be seen that for this submanifold there are no bifurcations when the atomic imbalance is non-zero, with bifurcations occuring at $\alpha=\pm 1$ when the atomic imbalance is zero. For the case when the atomic imbalance is non-zero, the global minimum of the classical Hamiltonian (\ref{ham2}) occurs when $\phi=N\pi/4$ and $z$ is the unique solution of (\ref{sol1}). In particular, for the solution $z\in[2k-1,1]$ $dz/d\alpha$ is a continuous function of $\alpha$.   
When the atomic imbalance is zero and $\alpha>1$, the global minimum of the classical Hamiltonian (\ref{ham2}) always occurs at the phase space boundary $z=-1$ with $\phi$ arbitrary. At $\alpha=1$ a bifurcation occurs, and for $\alpha$ slightly less than 1 two saddle points arise for $z=-1$ with $\phi$ given by solution to (\ref{ssol3}) and a new global minimum emerges corresponding to $\phi=N\pi/4$ with $z$ the unique solution of (\ref{ssol1}). In this case $dz/d\alpha$ is discontinuous at $\alpha=1$.       

In the following sections we will conduct an analysis of the quantum Hamiltonian (\ref{ham}). In particular we will establish that the bifurcation occuring at $(\alpha,\lambda)=(1,0)$ when the atomic imbalance is zero can be seen to influence the ground-state properties of the quantum system. In the context of the quantum system we will refer to the boundaries in Figs. \ref{curvekdiff}, 
\ref{parametro1} as  {\it threshold couplings}. We avoid using the terminology {\it quantum phase transition} as the analysis is conducted for finite particle number, not in the thermodynamic limit. The reason for not taking the thermodynamic limit is that the quantities $\lambda$ and $\alpha$ are dependent on $N$. Additionally, in the thermodynamic limit $N\rightarrow\infty$ with $k$ finite the semi-classical results predict qualitative differences between the cases $k=0$ and $k\neq 0$. However if $N$ is odd then we cannot have $k=0$, raising technical issues about whether the limit is convergent. Consequently we only consider the case of finite particle number. To deal with the subtleties of the finite size of the system we will formally define a quantum phase {\it pre-transition} in Sect. \ref{wo}.

%%%%%%%%%%%

%\end{document} 
 \section{Exact solution of the quantum Hamiltonian}

We now turn our attention to a quantum mechanical treatment of the model, to investigate the nature of the additional
threshold couplings when the atomic imbalance is zero. First we derive an exact Bethe ansatz solution of the model, and then use this to map the spectrum of the Hamiltonian (\ref{ham}) into the spectrum of a one-body Schr\"odinger operator.
 
\subsection{Energy eigenvalues as roots of a polynomial equation}

We rewrite the system Hamiltonian in a compact form as 
\begin{equation}
H=U+\Omega(a^{\dagger}b^{\dagger}c+c^{\dagger}ba)
\end{equation}
where the operator $U$ is a function of the number operators:
\begin{align*}
U&=U_{aa}N_{a}^{2}+U_{bb}N_{b}^{2}+U_{cc}N_{c}^{2}+U_{ab}N_{a}N_{b}+U_{ac}N_{a}N_{c}+U_{bc}N_{b}N_{c}\\
&\qquad+\mu_{a}N_{a}+\mu_{b}N_{b}+\mu_cN_c.
\end{align*}
Since the operators $N$ and $k=J/N$ are conserved we fix these and without loss of generality consider cases where 
$k\geq 0$. This restricts the Hilbert space to a subspace of dimension $(m+1)$ spanned by the vectors 
\begin{equation}
 \left|{l-j};{m-j};{j}\right>
\label{basis}
\end{equation}
where we have defined 
\begin{align}
l=\frac{N(1+k)}{2},\nonumber\\
m=\frac{N(1-k)}{2}\nonumber
\end{align}
such that $l+m=N$. 
We then look for eigenstates of (\ref{ham}) of the form 
\begin{equation}
\label{eq:gstate}
|\psi\ket=\sum_{j=0}^{m}\rho_j \left|{l-j};{m-j};{j}\right>
\end{equation}

Since the basis states (\ref{basis}) are eigenstates of each of the number operators they are eigenstates
 of the operator $U$ so we can define the quantities ${\mathcal U}_j$ through
\begin{equation}
U \left|{l-j};{m-j};{j}\right>
={\mathcal U}_j \left|{l-j};{m-j};{j}\right>.
\end{equation}
The Hamiltonian acts on the general state (\ref{eq:gstate}) as 
\begin{align}
H |\psi\ket &=\sum_{j=1}^{m-1} ({\mathcal U}_{j}\rho_j + \Omega((j+1)\rho_{j+1} + (l+1-j)(m+1-j)\rho_{j-1}))  
\left|l-j;m-j;j\right> \nonumber\\
& \qquad +({\mathcal U}_0 \rho_0 + \Omega\rho_1)\left|l;m;0\right> + 
({\mathcal U}_m \rho_m + \Omega \rho_{m-1}(l-m+1)) \left|l-m;0;m\right> \label{Hside}
\end{align}
Requiring that (\ref{eq:gstate}) is an eigenstate of the Hamiltonian with energy eigenvalue $E$
leads to the following recursion relations that must be satisfied by coefficients $\rho_j$: 
\begin{subequations}
\label{recrels}
\begin{align}
\Omega\rho_1+{\mathcal U}_0\rho_0&=E\rho_0,\label{rho0} \\
\Omega((j+1)\rho_{j+1}+(l+1-j)(m+1-j)\rho_{j-1})+{\mathcal U}_j\rho_j&=E\rho_j, 
\label{rhoj} \\
{\mathcal U}_m\rho_m +\Omega(l-m+1)\rho_{m-1} &= E \rho_m, \label{rhom} 
\end{align}
\end{subequations}
where $1<j<m-1$ in (\ref{rhoj}). 
As the normalisation of the state (\ref{eq:gstate}) can be chosen arbitrarily, we have the freedom to choose 
$\rho_0=1$. The recursion relation (\ref{rhoj}) then shows that $\rho_j$ is a polynomial in $E$ of order $j$. 
The constraint (\ref{rhom}) is thus a polynomial in $E$ of order $(m+1)$, whose roots are the energy eigenvalues of 
(\ref{ham}). Since the number of roots is the same as the dimension of the subspace spanned by the vectors (\ref{basis}), 
all energy eigenvalues are given by the roots of (\ref{rhom}).

\subsection{Bethe ansatz solution and mapping to a Schr\"odinger equation}

With the above implicit form for the energy eigenvalues we are able to map the energy spectrum into that
 of a one-dimensional Schr\"odinger equation.  We start by mapping the energy eigenstates to polynomial solutions
of a particular second-order ordinary differential equation (ODE) and then utilise a change of variables such that the differential equation takes the form of the
 Schr\"odinger equation. 
Each eigenstate of the system (\ref{eq:gstate}) can be represented by an $m^{\text{th}}$ order polynomial with coefficients $\rho_j$ ($j=1,2,..m.$).  For a particular energy, we can then construct an ODE for $G(u)$ such that the polynomial coefficients must satisfy the recursion relations of (\ref{recrels}). Below we outline the details of this construction.  

Consider a general second-order ODE eigenvalue problem satisfied by an $m^{\text{th}}$ polynomial function $G(u)$:
\begin{equation}
a(u)G^{''}+b(u)G'+c(u)G=EG
\label{ODEform}
\end{equation}
First we write the polynomial $G(u)$ with roots $\{u_p\}_{p=1}^m$ in the factorised form 
\begin{equation*}
G(u)=\prod_{p=1}^{m}(u-u_{p})
\end{equation*}
such that 
\begin{eqnarray*}
G'(u)&=&\sum_{p=1}^m\prod_{q\neq p}^m(u-u_q), \\ 
G''(u)&=&\sum_{p=1}^m\sum^m_{q\neq p}\prod^m_{\substack{r\neq p\\r\neq q}}(u-u_r).
\end{eqnarray*}
Evaluating (\ref{ODEform}) at the root $u_q$ leads to the Bethe ansatz equations  
\begin{equation}
\label{BAeqns}
\frac{b(u_q)}{a(u_q)}=\sum_{p\neq q}^{m}\frac{2}{u_p-u_q},\hspace{1cm}q=1,2,...,m.
\end{equation}
Hence, the roots of the polynomial must satisfy the system of coupled equations (\ref{BAeqns}) if $G(u)$ is a solution to (\ref{ODEform}).

We can map the  solutions of (\ref{ODEform}) with eigenvalue $E$ to solutions of a Schr\"odinger equation 
\begin{equation}
\label{shro}
\frac{-d^2\psi(x)}{dx^2}+V(x)\psi(x)=E\psi(x)
\end{equation}
with the same eigenvalues,
by mapping the polynomial solution of (\ref{ODEform}) to a wavefunction of (\ref{shro}) via 
\begin{equation*}
\psi(x)=e^{f(x)}{G(u(x))}.
\end{equation*}
Substituting into the Schr\"odinger equation gives the following relations to be satisfied
\begin{subequations}
\label{mapp}
\begin{align}
a(u)&=-\left(\frac{du}{dx}\right)^2 \label{a}\\
b(u)&=-\frac{d^2u}{dx^2}-2\frac{du}{dx}\frac{df}{dx}\label{b}\\
c(u)&=V(x)-\frac{d^2f}{dx^2}-\left(\frac{df}{dx}\right)^2\label{c}
\end{align}
\end{subequations}

In view of the above discussions, we now formulate the Bethe ansatz solution for (\ref{ham}) and the associated 
mapping to a Schr\"odinger equation.
To simplify the notation, we define
\begin{equation*}
{\mathcal U}_j = A(m-j)(m-j-1)+ B(m-j)+ C
\end{equation*}
where 
\begin{align*}
A&=U_{aa} +U_{bb} +U_{cc} +U_{ab}-U_{ac}-U_{bc}\\
B&=(1+2l-2m)U_{aa}+U_{bb} +(1-2m)U_{cc} +(1+l-m)U_{ab}\\
&~~~~~~+(2m-l-1)U_{ac} +(m-1)U_{bc}+\mu_a+\mu_b-\mu_c \\
C&=(1-m)^2U_{aa}+m(l-m)U_{ac} +m^2 U_{cc}+(m-l)\mu_a +m \mu_c.\\
\end{align*}
The polynomial defined as
\begin{equation}
G(u)=\sum_{j=0}^{m}\rho_{j}u^{m-j},
\end{equation}
with the $\rho_j$ satisfying (\ref{rho0},\ref{rhoj},\ref{rhom}), is a solution to the following differential equation
\begin{equation}
(Au^2+\Omega u)G^{''}+(Bu+\Omega(l-m+1-u^2))G^{'}+(\Omega mu+C)G=EG.
\label{ODEG}
\end{equation}
The roots of $G(u)$ are solutions of the Bethe ansatz equations
\begin{equation}
\frac{\Omega(l-m+1-u^2_q)+Bu_q}{u_q(\Omega+Au_q)}=\sum^m_{p\neq q}\frac{2}{u_p-u_q},\hspace{1cm}q=1,2,...,m.
\end{equation}
We can also derive an expression for the energy eigenvalues of the model in terms of the roots $u_q$. Consider the leading order expansions
\begin{align*}
G(u)&=u^m- u^{m-1} \sum_{q=1}^{m} u_{q} +...\\
G'(u)&=m u^{m-1}-(m-1)u^{m-2}\sum_{q=1}^m u_{q}+...\\
G''(u)&=m(m-1)u^{m-2}-(m-1)(m-2)u^{m-3} \sum_{q=1}^m u_{q} +...\\
\end{align*}
We substitute these expressions into (\ref{ODEG}) and equate terms of order $m$ to arrive at the following expression for the energy eigenvalues of the system
\begin{equation}
E=A m(m-1)+ Bm+C-\Omega \sum_{q=1}^{m}u_{q}
\end{equation}

Next we determine the explicit form of the Schr\"odinger equation. 
Comparing (\ref{ODEG}) to  (\ref{ODEform}) gives
\begin{align*}
a(u)&=Au^2 + \Omega u\\
b(u)&=(l-m+1-u^2)\Omega+Bu\\
c(u)&=mu\Omega+C
\end{align*}
Using (\ref{a},\ref{b},\ref{c}) we may perform the mapping to the Schr\"odinger equation by choosing
\begin{align*}
\frac{du}{dx}&=\pm\sqrt{-Au^2-\Omega u}\\
\end{align*}
Integrating this expression (with a convenient choice for the constant of integration) gives
\begin{align}
u=\frac{\Omega}{2A}(\cos({\sqrt{A}x})-1)
\end{align}
We also find that
\begin{align*}
\frac{df}{dx}&=\frac{\Omega^2}{4A^{\frac{3}{2}}}\sin({\sqrt{A}x})+\left(\sqrt{A}(l-m+1)-\frac{B}{2\sqrt{A}}-\frac{\Omega^2}{2A^{3/2}}\right)\csc(\sqrt{A}x)\\
&\qquad+\left(\frac{-\sqrt{A}}{2}+\frac{\Omega^2}{2A^{{3}/{2}}}+\frac{B}{2\sqrt{A}}\right)\cot{(\sqrt{A}x)}
\end{align*}
So the wavefunction
\begin{equation}
\Psi(x)=\exp\left(f(x)\right)\prod_{p=1}^{m}\left(\frac{\Omega}{2A}(\cos({\sqrt{A}x})-1)-u_p\right)
\label{wf}
\end{equation}
satisfies the Schr\"odinger equation (\ref{shro}) with potential
\begin{align*}
V(x)&=mu\Omega+C+\frac{d^2f}{dx^2}+\left(\frac{df}{dx}\right)^2\\
&=\left(C+\frac{\Omega^2}{2A}(l-2m+2)-\frac{\Omega^4}{2A^3}-\frac{A}{4}
+B\left(\frac{1}{2}-\frac{3\Omega^2}{4A^2}-\frac{B}{4A}\right)\right)\\
&\quad+\frac{\Omega^4}{16A^3}\sin^2(\sqrt{A}x)+\frac{\Omega^2}{2A}\left(m+\frac{\Omega^2}{2A^2}+\frac{B}{2A}\right)
\cos({\sqrt{A}x})\\
&\quad+\left(\frac{3A}{4}+A(l-m+1)^2+\frac{\Omega^2}{A}(l-m+1)+\frac{\Omega^4}{2A^3}-\frac{\Omega^2}{A}
+B\left(\frac{B}{2A}+\frac{\Omega^2}{A^2}-1\right)\right) \\
&\qquad\qquad \times \csc^2({\sqrt{A}x})\\
&\quad+\left(\left(\frac{\Omega^2}{A}+B\right)(l-m+2)-2A(l-m+1)-\frac{\Omega^4}{2A^3}
-B\left(\frac{B}{2A}+\frac{\Omega^2}{A^2}\right)\right)\\
&\qquad\qquad \times \cot(\sqrt{A}x)\csc(\sqrt{A}x). 
\end{align*}
The above potential is an example of a quasi-exactly solvable potential \cite{quasi}, whereby a finite number of eigenstates of the form (\ref{wf}) can be constructed.  The concept of mapping the spectrum of many-body systems into those of one-body Schr\"odinger equations has been discussed in detail in \cite{uz}.
 
\section{Analysis in the no scattering limit}

In this section we now conduct a deeper analysis of the Hamiltonian in the no scattering limit where 
$U_{jk}=0$ for all $j,k=a,b,c.$ In this limit the model simplifies substantially, yet remains sufficiently non-trivial to enable us to gain an understanding of the quantum behaviour through the Schr\"odinger equation mapping, ground-state expectation values and quantum dynamics. Specifically the no scattering limit corresponds to the coupling $\lambda=0$ in the semi-classical analysis of Sect. \ref{sca}. With reference to Fig. \ref{parametro1} there are two threshold couplings in the case of zero atomic imbalance. One occurs at $(\alpha,\lambda)=(1,0)$, signifying the bifurcation of the global minimum of the Hamiltonian, while the other occurs at 
$(\alpha,\lambda)=(-1,0)$, signifying the bifurcation of the global maximum. In contrast there are no bifurcations along the line $\lambda=0$ in Fig. \ref{curvekdiff}. We focus our attention to the coupling $(\alpha,\lambda)=(1,0)$ as the bifurcation of the fixed point in phase space is associated with the ground state of the quantum system.   

\subsection{Schr\"odinger equation mapping} \label{sem}

For small values of $A$, we can take series expansions for the trigonometric functions in the potential $V(x)$ and wavefunction $\Psi(x)$.
Then taking the limit $A\rightarrow 0$ (corresponding to $\lambda=0$ for the analogous classical system (\ref{ham2})) the Schr\"odinger potential becomes
\begin{eqnarray}
V(x)&=&C-\frac{B}{2}(N+1)+ \left(J^2-\frac{1}{4}\right)x^{-2} \nonumber \\
&&\qquad + \left( \frac{B^2}{16} - \frac{ \Omega^2}{8}(N+2) \right)x^2 
+ \frac{\Omega^2B}{32}x^4 +\frac{\Omega^4}{256}x^6
\label{semigeneralpotential}
\end{eqnarray}
where we now parametrise the system in terms of the variable $J=l-m$ and $N=l+m$.
Now consider a simple subclass of the general Hamiltonian (\ref{ham})
\begin{equation}
H=\mu N_c +\Omega(a^{\dagger}b^{\dagger}c+c^{\dagger}ba)
\label{hamres}
\end{equation}
We have mapped the general model to a Schr\"odinger equation in the previous section.  
The above case (\ref{hamres}) corresponds to $A=0$, $B=-\mu$ and $C=m\mu$ in equation (\ref{semigeneralpotential}). The energy eigenstates map to solutions of the Schr\"odinger equation with potential
\begin{align}
\label{genpot}
V(x)&=\frac{\mu(N+1)}{2}+ \left(J^2-\frac{1}{4}\right) x^{-2}\nonumber\\
&\qquad + \frac{1}{16}\left(\mu^2-2\Omega^2 (N+2)\right) x^2 -\frac{\mu \Omega^2}{32} x^4+ \frac{\Omega^4}{256} x^6.
\end{align}
The associated wavefunction is given by
\begin{equation}
\Psi(x)=x^{(J+1/2)}\exp\left(\frac{-\Omega^2 x^4}{64}+\frac{\mu x^2}{8}\right)\prod_{p=1}^{m}\left(\frac{-\Omega x^2}{4}-u_p\right)
\label{wf1}
\end{equation}
with energy eigenvalues 
\begin{equation}
E= -\Omega \sum_{q=1}^{(N-J)/2}u_{q}
\end{equation}
where the $\{u_q\}$ are solutions to the Bethe ansatz equations 
\begin{equation}
\frac{(J+1)}{u_q}-u_q-\frac{\mu}{\Omega}=\sum_{p\neq q}^{(N-J)/2}\frac{2}{u_p-u_q},\hspace{1cm}q=1,2,...,(N-J)/2.
\end{equation}
\begin{figure}[!h]
\begin{center}
\begin{tabular}{ccc}
\includegraphics[width=7cm,height=7cm]{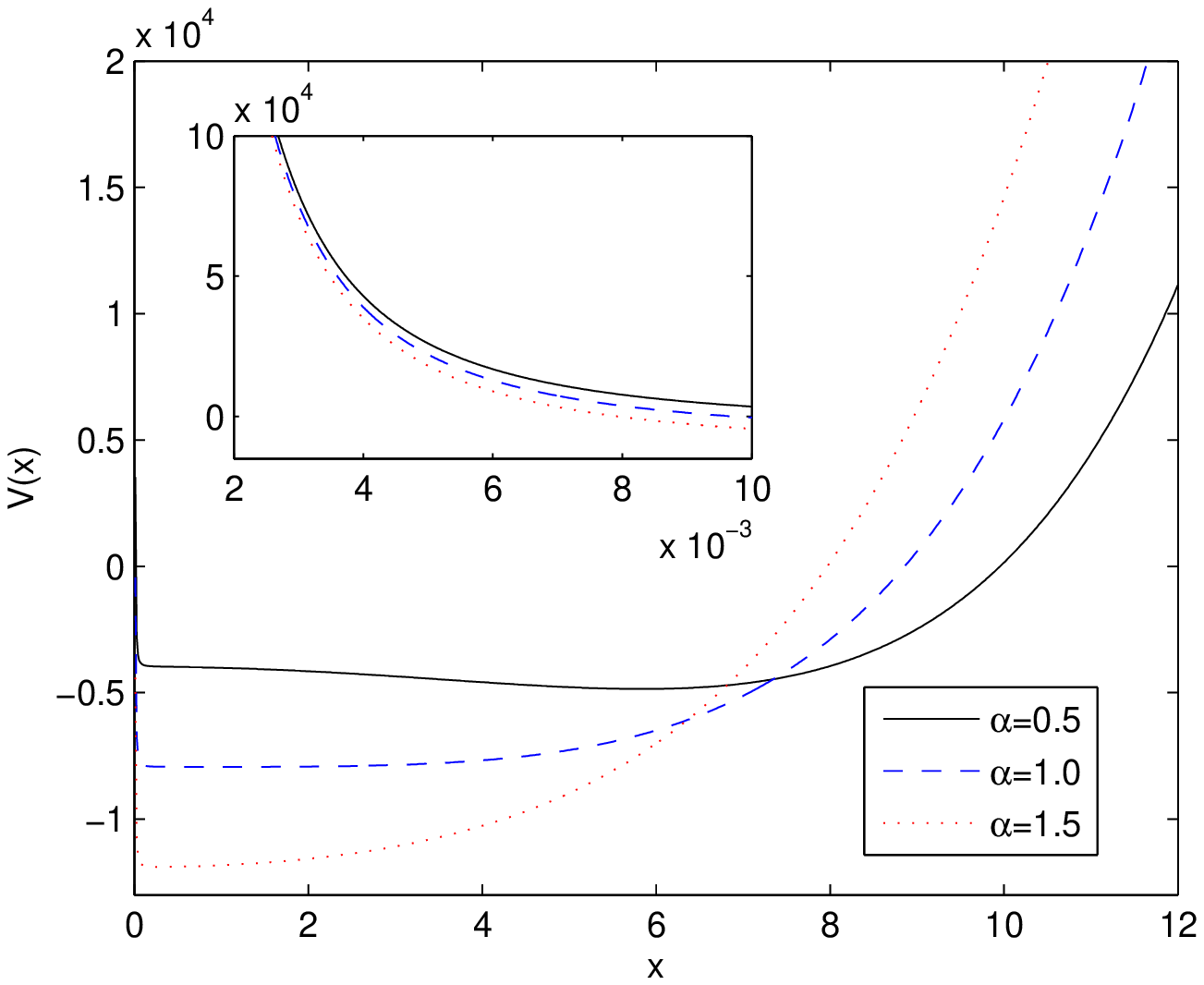}
    &\quad &
\includegraphics[width=7cm,height=7cm]{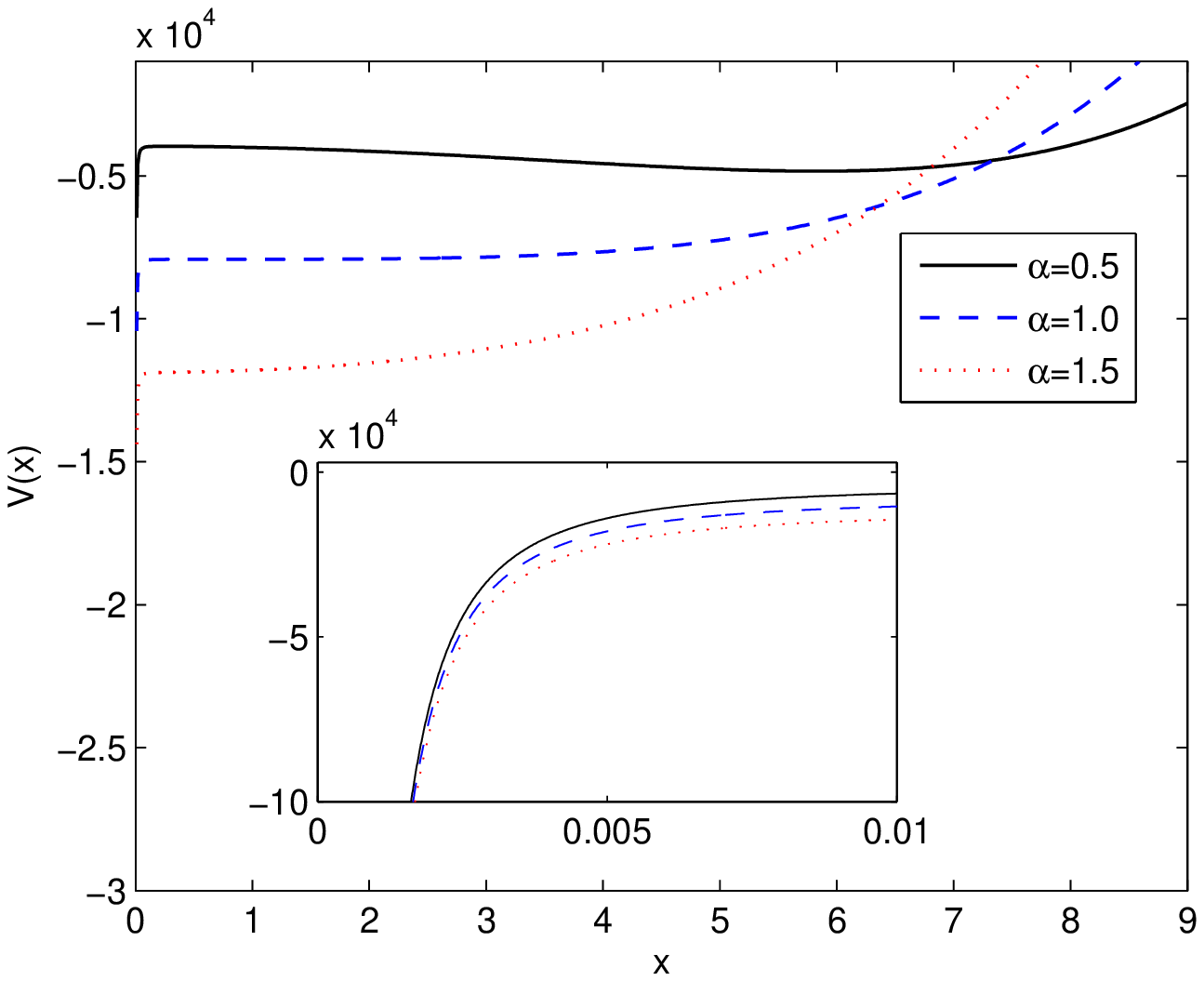}
  \\
 (a)  &\quad & (b)  \\     
\end{tabular}     
\end{center}
\caption{ The potential $V(x)$ given by (\ref{genpot}). (a) $N=501$ and atomic imbalance $J=1$. The potential is bounded from below, with the inset showing  
$V\rightarrow\infty $ as $x\rightarrow 0$. Varying the coupling parameter $\alpha$ across the threshold value 
$\alpha=1$, it is apparent there is no bifurcation of the potential minimum. (b) $N=500$ and atomic imbalance $J=0$. The potential is not bounded below, with the inset showing $V\rightarrow-\infty $ as $x\rightarrow 0$.  
As the coupling parameter $\alpha$ is varied across the threshold value $\alpha=1$, it is apparent there is a bifurcation with the formation of a local minimum and a local maximum for $\alpha<1$. }
\label{potentials}
\end{figure}

The semi-classical analysis predicts a threshold coupling at $\alpha=1$ when the atomic imbalance is zero where for the case under consideration 
$\alpha=-\mu/(\Omega\sqrt{2N})$. When the atomic imbalance is non-zero there is no predicted threshold coupling. 
Fig. \ref{potentials} (a) depicts the potential (\ref{genpot}) for $N=501,\, J=1$ and various values of $\alpha$ close to the threshold value $\alpha=1$. 
It can be seen that the potential has a single minimum for all $\alpha$. In contrast, Fig. \ref{potentials} (b) shows (\ref{genpot}) for $N=500$ and $J=0$. For this case the potential is not bounded from below and there is a bifurcation for $\alpha\approx 1$. For the model (\ref{hamres}), the predictions of the semi-classical analysis conducted in Sect. \ref{sca} of a threshold coupling at $\alpha\approx 1$ are consistent with qualitative differences of the associated Sch\"odinger equation.  

Now we examine bifurcations of the critical points of the potential (\ref{genpot}) in more detail. Consider the general class of potentials 
\begin{eqnarray}
V(x)&=& \A x^{-2}+\B x^2+\C x^4 +\D x^6 \label{generic} 
\end{eqnarray}
where $\C,\,\D$ are assumed to be positive and no constraints are placed on $\A$ nor $\B$. In particular we wish to determine when the condition
\begin{eqnarray}
\frac{dV}{dx}=\frac{d^2V}{dx^2}=0  \label{condition}
\end{eqnarray} 
can be met. Since the potential is a symmetric function, we restrict to non-negative values of $x$. When $\A=0$ it is straightforward to deduce that, for any values of $\C$ and $\D$,
 (\ref{condition}) holds at $x=0$ when $\B=0$. 
For non-zero values of $\A$ we find      
\begin{eqnarray}
\frac{dV}{dx}&=& -2\A x^{-3}+2\B x +4\C x^3 +6\D x^5 \label{dvdx} \\
\frac{d^2V}{dx^2}&=& 6\A x^{-4} +2\B +12 \C x^2 +30\D x^4 \label{d2vdx2} 
\end{eqnarray}
We set both (\ref{dvdx}) and (\ref{d2vdx2}) to zero and take particular linear combinations to obtain the following relations:
\begin{eqnarray}
\frac{1}{8}\frac{d^2V}{dx^2}+\frac{3}{8x}\frac{dV}{dx}&=& \B  + 3 \C x^2  +6 \D x^4 =0\label{bcd} \\
\frac{x^4}{8}\frac{d^2V}{dx^2}-\frac{5x^3}{8}\frac{dV}{dx}&=& 2 \A  - \B x^4 -\C x^6=0 \label{abc} 
\end{eqnarray}

Note that eq. (\ref{bcd}) is independent of $\A$, and has solutions
\begin{eqnarray}
x^2&=&\frac{-3\C\pm\sqrt{9\C^2-24\B\D}}{12\D}. \label{x}
\end{eqnarray}
We take the positive square root in (\ref{x}) and impose $\B<0$, ensuring $x^2>0$.  
Treating $\B$ as a small parameter such that 
\begin{eqnarray}
\left|\B\right|\ll  \frac{\C^2}{\D} \label{muchless}
\end{eqnarray} 
yields 
\begin{eqnarray}
x^2\approx-\frac{\B}{3\C}. \label{z}
\end{eqnarray} 
Next we substitute (\ref{z}) into (\ref{abc}) and solve for $\B$:
\begin{equation}
\B=3(\A\C^2)^{1/3}. \label{B}
\end{equation}
Since $\B$ is negative such a solution only exists when $\A$ is negative. 

Matching the co-efficients between (\ref{genpot}) and (\ref{generic}) gives 
\begin{eqnarray*}
\A&=&  J^2-\frac{1}{4} \\
\B&=&  \frac{1}{16}\left(\mu^2-2\Omega^2(N+2)\right)  \\
\C&=&   -\frac{\mu \Omega^2}{32}\\
\D&=& \frac{\Omega^4}{256} 
\end{eqnarray*}
When $\alpha\approx 1$, or equivalently $\mu\approx -\Omega\sqrt{2N}$, we satisfy the requirements $\C,\,\D>0$ and 
(\ref{muchless}). A bifurcation of the potential only occurs when the atomic imbalance $J$ is zero, as $\A$ must be negative. Using 
(\ref{B}) we then find $\mu$ satisfies 
$$3\mu^{2/3}\Omega^{4/3}+\mu^2=2\Omega^2(N+2). $$ 
From this expression we determine the leading quantum correction to the semi-classical result for the threshold coupling
for (\ref{hamres}):   
$$\mu\approx-\Omega (2(N+2))^{1/2}+\frac{3\Omega}{2}(2(N+2))^{-1/6}. $$

\subsection{Ground-state expectation values and quantum dynamics} 

\begin{figure}[ht]
\begin{center}
\epsfig{file=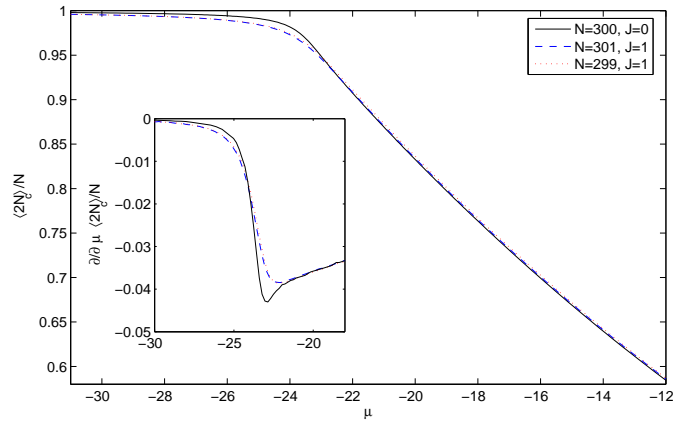,scale=0.45} %width=12cm,height=7cm,angle=0}
\end{center}
\caption{Ground-state expectation values of the (scaled) molecular number operator $N_c$ as a function of the coupling $\mu$, for the Hamiltonian
(\ref{hamres}). Results shown correspond to $\Omega=1$ and both zero and non-zero atomic imbalance. The inset shows the first derivative of the expectation values with respect to the coupling $\mu$. While there are quantitative differences there is no significant qualitative change between the case of zero and minimal non-zero atomic imbalance.}  
\label{expectationvalue}
\end{figure} 

Next we examine the behaviour of the ground state of (\ref{hamres}) as the threshold coupling $\alpha=1$ is crossed. From the semi-classical analysis we have found that the global minimum for $\alpha>1$ and $J=0$ occurs at $z=-1$ in phase space. In the Hilbert space of states this corresponds to $|0;0;N/2  \rangle$. It is then appropriate to compute 
the gound-state expectation value $\langle N_c\rangle$ for the quantum system as the coupling is varied. Results are shown in Fig. \ref{expectationvalue}. In general agreement with the semi-classical result, it can be seen that the expectation value $\langle N_c\rangle/N$ is close to unity 
when $-\mu>\Omega\sqrt{2N}$ for the case of zero atomic imbalance. When $-\mu<\Omega\sqrt{2N}$ the expectation value decreases. The figure also shows that when the imbalance $J=1$ the results are qualitatively similar. However from the predictions of both the semi-classical analysis of Sect. \ref{sca} and the the associated one-body Schr\"odinger potential of Sect. \ref{sem} we do not obtain any prediction about the change in the ground-state properties when the imbalance is non-zero. Further increase in the value of $J$ (not shown) does not indicate any dramatic change in the qualitative features of $\left<2 N_c\right>/N$ as a function of $J$. As mentioned previously, because we are studying a finite system changes in the ground-state properties are smooth as $J$ is varied. If we instead look at the quantum dynamics as the threshold coupling is crossed, qualitative differences are more apparent.  

\begin{figure}[ht]
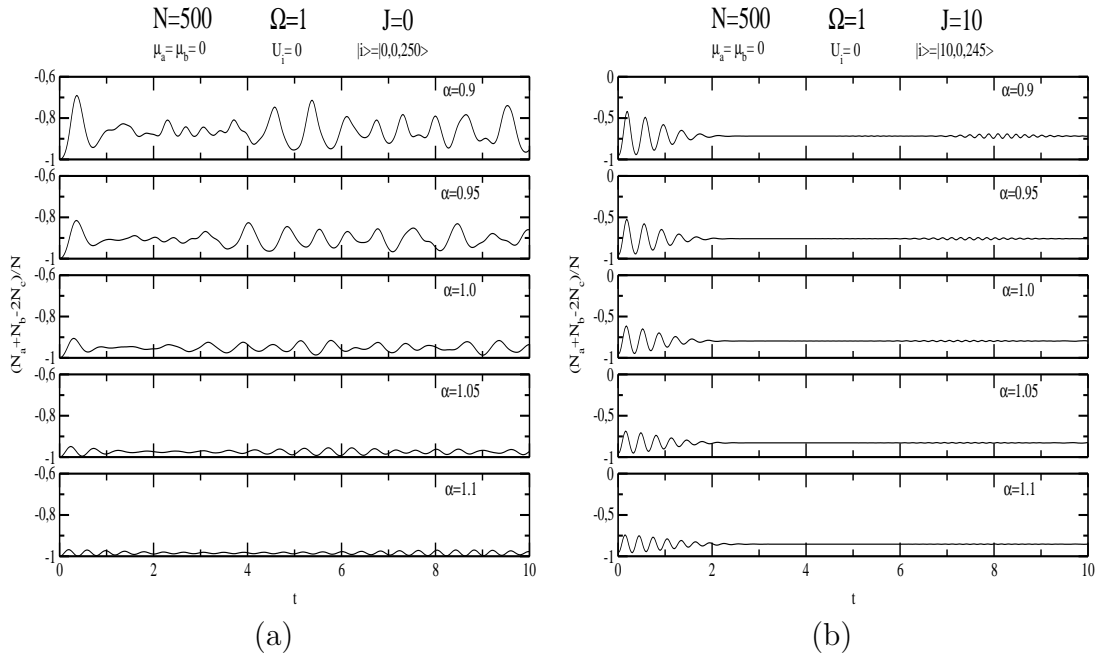

\begin{center}
\begin{tabular}{cc}
\epsfig{file=figures/Fig7a.eps,width=7cm,height=8cm,angle=0}    &
\epsfig{file=figures/Fig7b.eps,width=7cm,height=8cm,angle=0}   \\
 (a)  & (b)  \\     
\end{tabular}     
\end{center}
\caption{Time evolution of the expectation value of $z$ for the Hamiltonian (\ref{hamres}) with $N=500$. The cases shown are, from top to bottom, $\alpha=0.9,\,0.95,\,1,\,1.05,\,1.1$. (a) $J=0$ and initial state $|0;0;250\rangle$. The oscillations are largely irregular with significantly decreasing amplitude as the point at $\alpha=1$ is crossed. This point corresponds to the boundary at $(\alpha,\lambda)=(1,0)$ between regions $III$ and $IV$ as shown in Fig. \ref{parametro1}.  
(b) $J=10$ with initial state 
$|10;0;245\rangle$. The oscillations display collapse and revival behaviour with smoothly decreasing amplitude as the point at $\alpha=1$ is crossed, indicative of the fact there is no boundary at $(\alpha,\lambda)=(1,0)$ in Fig. \ref{curvekdiff}.} 
\label{quantumdynamics}
\end{figure} 

In general the time evolution of any state is given 
by $|\Psi(t) \rangle = U(t)|\phi \rangle$, 
where $U(t)$ is the temporal evolution operator $\displaystyle U(t)=\sum_{j=0}^{m}|j\rangle \langle j|\exp(-i E_{j} t)$,  
$|j\rangle$ is an eigenstate with energy $E_{j}$ and $|\phi \rangle$ represents 
the initial state with $N=N_a+N_b+2N_c$.
We adopt the method of directly diagonalising the Hamiltonian (\ref{hamres}) as done in \cite{fitonel},  
and compute the expectation value of $z(t)$ through
$$
%\begin{equation}
\langle z(t)\rangle=\frac{1}{N}\langle \Psi (t)|N_a+N_b-2N_c|\Psi (t)\rangle.
%\end{equation}
$$
For a fixed atomic imbalance $J$ we will use the initial state configuration $|J;0;(N-J)/2  \rangle$. When $J=0$ this state correspond to $z=-1$ in phase space, which is a fixed point when $\alpha>1$. We thus expect that in this case 
$\langle z(t)\rangle$ will not vary significantly in time (i.e. the system is {\it localised}). 
On the other hand when $J\neq 0$ the state 
$|J;0;(N-J)/2  \rangle$ does not correspond to a fixed point.  
We therefore compare  the two cases of the quantum dynamics, with atomic imbalance $J=0$ and $J\neq 0$, as the value $\alpha=1$ is crossed. 
We fix the parameter $\Omega=1$ and use $\mu$ as the variable coupling parameter.

Results of the expectation value for $z$ are shown in Fig. \ref{quantumdynamics} for the cases of zero and non-zero atomic imbalance. The qualitative difference are quite apparent. In the case of zero atomic imbalance ($k=0$), Fig. \ref{quantumdynamics} (a), we find that for $\alpha<1$ there are irregular oscillations in $z$. By comparison the dynamics in Fig. \ref{quantumdynamics} (b) for non-zero imbalance ($k=0.02$) show a collapse and revival of oscillations. As the coupling parameter $\alpha$ is increased across the threshold value at $\alpha=1$, the transition to localised oscillations is much sharper in case (a) compared to case (b). Note in particular the vertical scales in (a) and (b) are not the same.  
We remark that the nature of the dynamics for $\alpha>1$ {\it does} change smoothly from localisation to delocalisation over the intermediate values $0<k<0.02$ (not shown). 
Taking the thermodynamic limit $N\rightarrow \infty$ does not aid in the analysis. For the Hamiltonian (\ref{hamres}) the condition for localisation of oscillations for zero atomic imbalance, $\alpha> 1$, is equivalent to $-\mu>\Omega\sqrt{2N}$. Hence for fixed $-\mu>0$ and $\Omega>0$ this condition imposes an upper bound on $N$ for which localisation occurs.

\section{Wavefunction overlaps} \label{wo}
\begin{figure}[ht]
\begin{center}
\epsfig{file=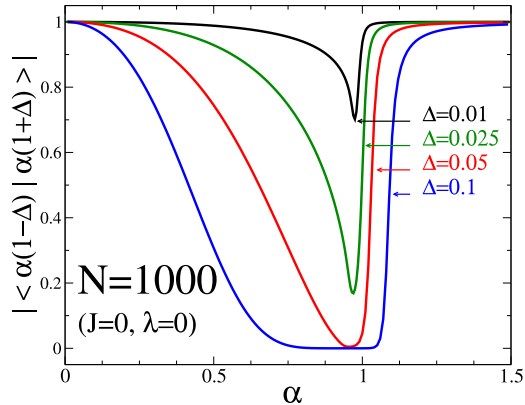,scale=0.25,angle=-90}          
\end{center}
\caption{Ground-state wavefunction overlaps for the Hamiltonian (\ref{hamres}) with $N=1000$ and various values of $\Delta$. The value of the local minimum at $\alpha\approx 1$ is a decreasing function of $\Delta$, asymptotically approaching zero.}
\label{gsoverlaps1}
\end{figure} 

In order to gain a better insight into the effect of the threshold couplings for the quantum system, in our final analysis we adopt the method of wavefunction overlaps \cite{zanardi,huanandjp}. 
If a system admits a quantum phase transition, then two states belonging to different phases of the same system are distinguishable. If states are distinguishable they must be orthogonal~\cite{nc} and consequently the wave function overlaps vanish. For systems which exhibit a quantum phase transition in the thermodynamic limit, the wavefunction overlaps between states in different phases go to zero in this limit. The occurrence of a minimum in the incremental ground-state wavefunction overlap in a finite system is then a precursor for a quantum phase transition in the thermodynamic limit. Thus for finite systems we identify quantum phase {\it pre-transitions} at couplings for which the incremental wavefunction overlap is (locally) minimal.  

\begin{figure}[ht]
\begin{center}
\begin{tabular}{ccc}
\epsfig{file=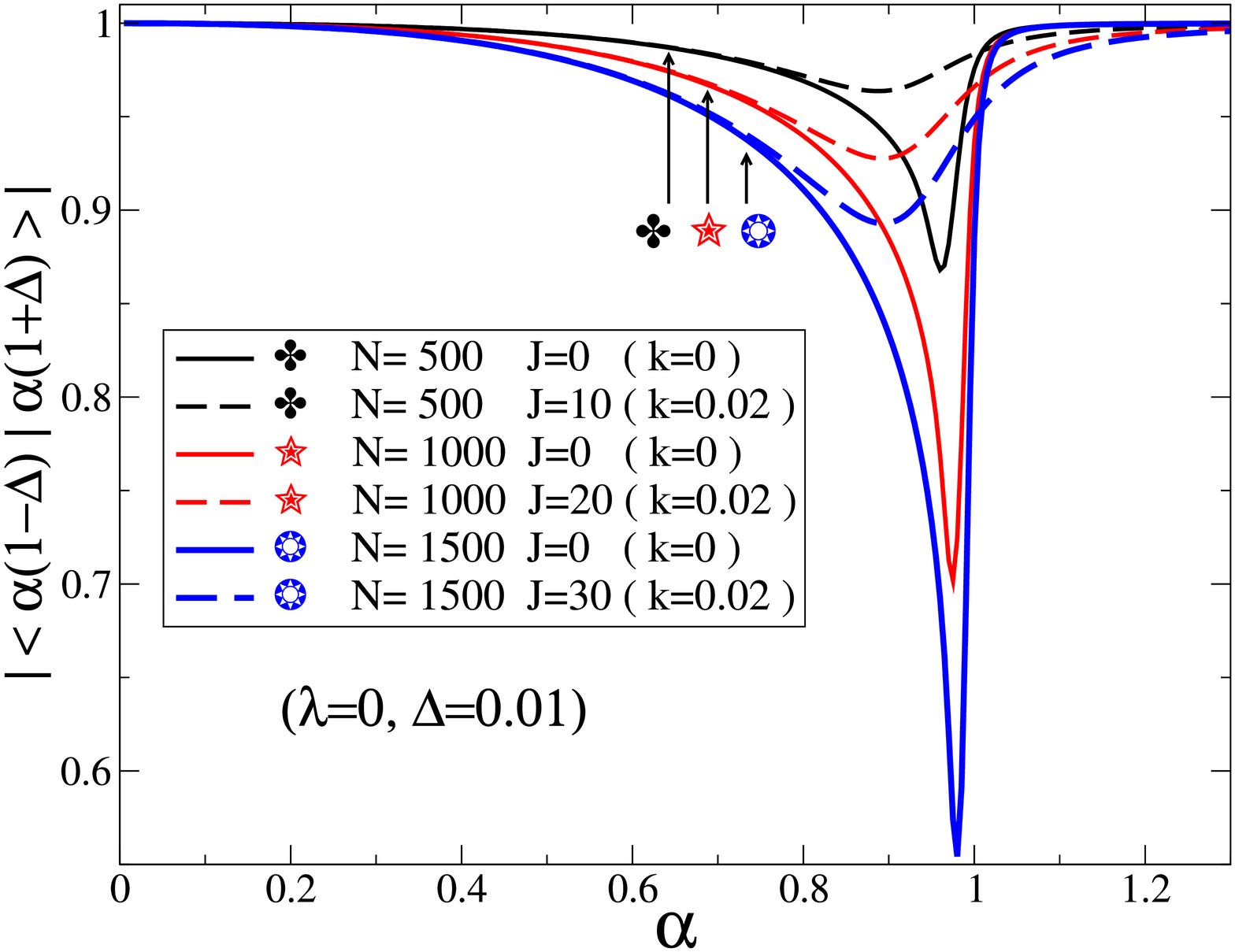,scale=0.25,angle=-90} &\quad&
\epsfig{file=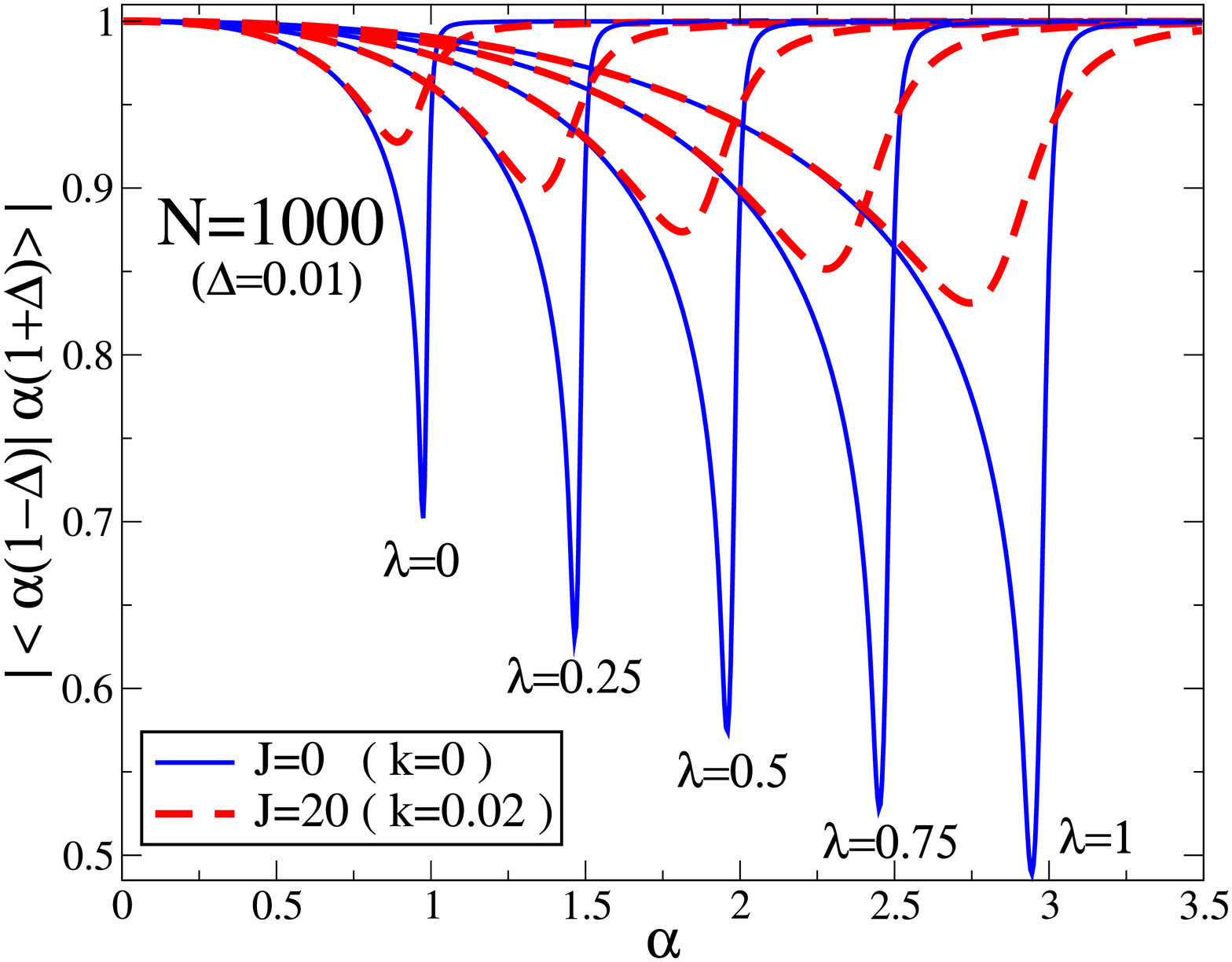,scale=0.25,angle=-90}          \\
(a)  &\quad & (b)  \\     
\end{tabular}          
\end{center}
\caption{ (a) Ground-state wavefunction overlaps of the Hamiltonian (\ref{hamres}), for $N=500,1000,1500$. The solid lines correspond to cases when the atomic imbalance is zero, while the dashed lines illustrate the behaviour for the fractional imbalance 
$k=0.02$. Two general properties that can be observed at the pre-transition coupling $\alpha\approx 1$ are (i) the minimum value decreases with increasing $N$; (ii) for fixed $N$, the value of the minimum is lower for $k=0$ compared to $k\neq 0$.
(b) Ground-state wavefunction overlaps of the Hamiltonian (\ref{hamres}) for $N=1000$ and different values of $\lambda$. 
The solid lines correspond to cases when the atomic imbalance is zero, while the dashed lines illustrate the behaviour for the fractional imbalance $k=0.02$. The locations of the minima fit the line of threshold couplings given by $\lambda=(\alpha-1)/2$ as predicted by the semi-classical analysis.}
\label{gsoverlaps2}
\end{figure} 

We now formally define   
a quantum phase pre-transition in terms of ground-state wavefunction overlaps. Let $H(\delta)$ denote a generic Hamiltonian depending on a coupling parameter $\delta$. Assuming the ground state of the system is non-degenerate, 
let $\left|\Psi(\delta)\right>$ denote the unique normalised ground state. For fixed small $\Delta$ we define the function 
$W_{\Delta}(\delta)  $ by 
\begin{eqnarray*}
W_{\Delta}(\delta)  = \left|\left<\Psi(\delta(1-\Delta))|\Psi(\delta(1+\Delta))\right>\right|
\end{eqnarray*}
which is symmetric in $\Delta$, bounded between 0 and 1, and satisfies $W_0(\delta)=1$. 
Generically, $W_{\Delta}(\delta)$ is a decreasing function of $\Delta$.  Fig. \ref{gsoverlaps1} shows the behaviour of the wavefunction overlaps for the Hamiltonian (\ref{hamres}) with $N=1000$, and different values of $\Delta$. It is clear that there is a distinct dip in the quantity $W_{\Delta}(\alpha)$ near the threshold coupling $\alpha=1$. The choice for 
$\Delta$ affects the magnitude of the minimum, which can be made arbitrarily small. In Fig. \ref{gsoverlaps1} when $\Delta=0.05$, representing a coupling change of about 
$5\%$ on either side of the threshold coupling, the ground states are essentially orthogonal.  However the value of $\alpha$ at which the minimum occurs is largely independent of $\Delta$.   
For a given $\Delta$  we say that there is a quantum phase pre-transition at $\delta_c$ if $W_{\Delta}(\delta)$, treated as a single-variable function of $\delta$, is locally minimal at $\delta_c$. 

We have computed the wavefunction overlaps for several cases with  both zero and non-zero atomic imbalance. Fig. \ref{gsoverlaps2} (a) shows the behaviour of $W_{\Delta}(\alpha)$ with $\lambda=0,\,\Delta=0.01$ and varying $N$ for both $k=0$ and $k=0.02$. It is clear the minimum value of $W_{\Delta}(\alpha)$, which determines the quantum phase pre-transition, is at $\alpha\approx 1$. The distinction between the predicted threshold coupling and the observed pre-transition coupling is that the pre-transition coupling also occurs for $k\neq 0$, although for fixed $N$ the value of the minimum is lower for 
$k=0$ compared to $k=0.02$. In all instances the value of the mimimum decreases with increasing $N$. Fig. \ref{gsoverlaps2} (b) shows similar results for fixed $N=1000$ and varying $\lambda$. In this latter case we see that the occurences of the mimima, determining the pre-transition couplings, fit well with the predicted boundary of threshold couplings given by $\lambda=(\alpha-1)/2$ (cf. Fig. \ref{parametro1}). However we again observe pre-transitions couplings for $k=0.02$ which were not predicted in Sect. \ref{sca}.   

In the above cases the minimum of $W_{\Delta}(\alpha)$ is substantially more pronounced for zero imbalance compared to non-zero imbalance. We can see that although quantum phase pre-transitions as defined exist for $k\neq 0$, the distiguishability of the two phases is more reliable in the limit as $k$ goes to zero. In the analyses of Sect. \ref{sca} and Sect. \ref{sem} qualitative differences are only found precisely when $k=0$. We interpret these results as the emergence of quantum phase boundaries at $k=0$.

\section{Discussion}

We have studied the ground-state phases of a three-mode model describing a heteronuclear molecular Bose--Einstein condensate, through a variety of techniques. Using a semi-classical analysis we were able to determine threshold couplings associated with fixed point bifurcations in phase space. We then derived the exact Bethe ansatz solution for the system and discussed how the spectrum of the Hamiltonian maps into that of an associated one-body Schr\"odinger equation. It was shown that in the particular subcase of no scattering interactions the threshold coupling for the global minimum in phase space for zero atomic imbalance was consistent with the existence of a bifurcation of the potential of the Schr\"odinger equation. For the non-zero imbalance case where the semi-classical results do not predict any threshold coupling, it was found that there was no bifurcation of the Schr\"odinger potential. These results suggested that the ground-state properties of the model are sensitive to whether the atomic imbalance was zero or non-zero. However due to the finite nature of the system, such a transition is not associated with a discontinuity that is defined in the thermodynamic limit.      
We then introduced the notion of a quantum phase pre-transition for finite systems, defined in terms of ground-state wavefunction overlaps. Applying this idea to the model under consideration we argued that a quantum phase boundary emerges in the limit as  the atomic imbalance goes to zero.       

For future work it would be useful to investigate the ground-state entanglement properties of the Hamiltonian. Not only would this be of interest in the study of the behaviour of the entanglement at the quantum phase boundary as $k$ approaches zero, but the model is also a simple example of a strictly tripartite system which may offer insights into the role of three-way entanglement in the description of quantum phases. 

\section*{Acknowledgements}

E.C.M. thanks CAPES-Coordena\c{c}\~ao de Aperfei\c{c}oamento de Pessoal de N\'{\i}vel Superior for financial support. 
A.T. and A.F. thank FAPERGS-Funda\c{c}\~ao de Amparo \`a Pesquisa do Estado do Rio Grande do Sul for financial 
support. A.F also acknowledges support from PRONEX under contract CNPq 66.2002/1998-99.
M.D., J.L. and N.O. are funded by the Australian Research Council through the Discovery Projects DP0557949 and DP0663773.

\vfil\eject

%\begin{appendix}
%\section{Appendix}
%\end{appendix}
\end{document}